\documentclass[10pt,twocolumn,letterpaper]{article}

\usepackage{cvpr}              
\definecolor{cvprblue}{rgb}{0.21,0.49,0.74}
\usepackage[pagebackref,breaklinks,colorlinks,allcolors=cvprblue]{hyperref}
\usepackage{multirow}
\usepackage{xcolor}
\usepackage{colortbl}     
\usepackage{algorithm}
\usepackage{algorithmic}
\usepackage{listings}
\definecolor{codegreen}{rgb}{0,0.6,0}
\definecolor{codegray}{rgb}{0.5,0.5,0.5}
\definecolor{codepurple}{rgb}{0.58,0,0.82}
\definecolor{backcolour}{rgb}{0.95,0.95,0.92}
\lstdefinestyle{mystyle}{
    backgroundcolor=\color{backcolour},   
    commentstyle=\color{codegreen},
    keywordstyle=\color{magenta},
    numberstyle=\tiny\color{codegray},
    stringstyle=\color{codepurple},
    basicstyle=\ttfamily\footnotesize,
    breakatwhitespace=false,         
    breaklines=true,                 
    captionpos=b,                    
    keepspaces=true,                 
    numbers=left,                    
    numbersep=5pt,                  
    showspaces=false,                
    showstringspaces=false,
    showtabs=false,                  
    tabsize=2
}

\newcommand{\best}[1]{\cellcolor{green!20}\textbf{#1}}
\newcommand{\secondbest}[1]{\cellcolor{yellow!20}#1}

\def\paperID{3777} 
\def\confName{CVPR}
\def\confYear{2026}

\title{LUMINA: A Multi-Vendor Mammography Benchmark with Energy Harmonization Protocol}

\author{Hongyi Pan$^{1\dagger}$, Gorkem Durak$^1$, Halil Ertugrul Aktas$^1$, Andrea M. Bejar$^1$, Baver Tutun$^2$,\\ Emre Uysal$^2$, Ezgi Bulbul$^2$, Mehmet Fatih Dogan$^2$, Berrin Erok$^2$, Berna Akkus Yildirim$^2$,\\ Sukru Mehmet Erturk$^3$, Ulas Bagci$^{1\dagger}$\\
\\
$^1$Department of Radiology, Northwestern University, Chicago, IL, USA\\
$^2$Department of Radiation Oncology, University of Health Sciences\\ Prof. Dr. Cemil Tascioglu City Hospital, Istanbul, Turkey\\
$^3$Department of Radiology, Istanbul University, Istanbul, Turkey\\
$^\dagger${\tt\small \{hongyi.pan, ulas.bagci\}@northwestern.edu}
}

\begin{document}
\maketitle

\begin{abstract}
Publicly available full-field digital mammography (FFDM) datasets remain limited in size, clinical annotations, and vendor diversity, hindering the development of robust models. We introduce \textbf{LUMINA}, a curated, multi-vendor FFDM dataset that explicitly encodes acquisition energy and vendor metadata to capture clinically relevant appearance variations often overlooked in existing benchmarks. This dataset contains 1824 images from 468 patients (960 benign, 864 malignant), with pathology-confirmed labels, BI-RADS assessments, and breast-density annotations. LUMINA spans six acquisition systems and includes both high- and low-energy imaging styles, enabling systematic analysis of vendor- and energy-induced domain shifts. To address these variations, we propose a \textit{foreground-only} pixel-space alignment method (``energy harmonization'') that maps images to a low-energy reference while preserving lesion morphology. We benchmark CNN and transformer models on three clinically relevant tasks: diagnosis (benign vs. malignant), BI-RADS classification, and density estimation. Two-view models consistently outperform single-view models. EfficientNet-B0 achieves an AUC of 93.54\% for diagnosis, while Swin-T achieves the best macro-AUC of 89.43\% for density prediction. Harmonization improves performance across architectures and produces more localized Grad-CAM responses. Overall, LUMINA provides (1) a vendor-diverse benchmark and (2) a model-agnostic harmonization framework for reliable and deployable mammography AI.
\end{abstract}

\section{Introduction}
Breast cancer is the most prevalent type of cancer and one of the leading causes of cancer-related deaths among women~\cite{lukasiewicz2021breast,wilkinson2022understanding}. 
Recent advances in deep learning have demonstrated the potential to improve diagnostic accuracy and detect subtle lesions that can be overlooked by human readers~\cite{shen2019deep,lotter2021robust}. However, training robust and generalizable models requires large-scale, high-quality datasets. In the mammography domain, publicly available datasets such as CBIS-DDSM~\cite{lee2017curated} and INbreast~\cite{moreira2012inbreast} are limited in size and heterogeneity, restricting the development of artificial intelligence (AI) systems that can perform reliably across diverse clinical settings. This underscores the need for comprehensive mammography datasets that capture high-resolution images across multiple views, patient populations, and imaging systems. To address this gap, we introduce the \textbf{LUMINA} \textbf{dataset }along with a foreground-only pixel-space CDF alignment method that reduces vendor/energy appearance drift and consistently improves accuracy and AUC across three tasks.

\begin{figure}[tb]
    \centering
    \includegraphics[width=.8\linewidth]{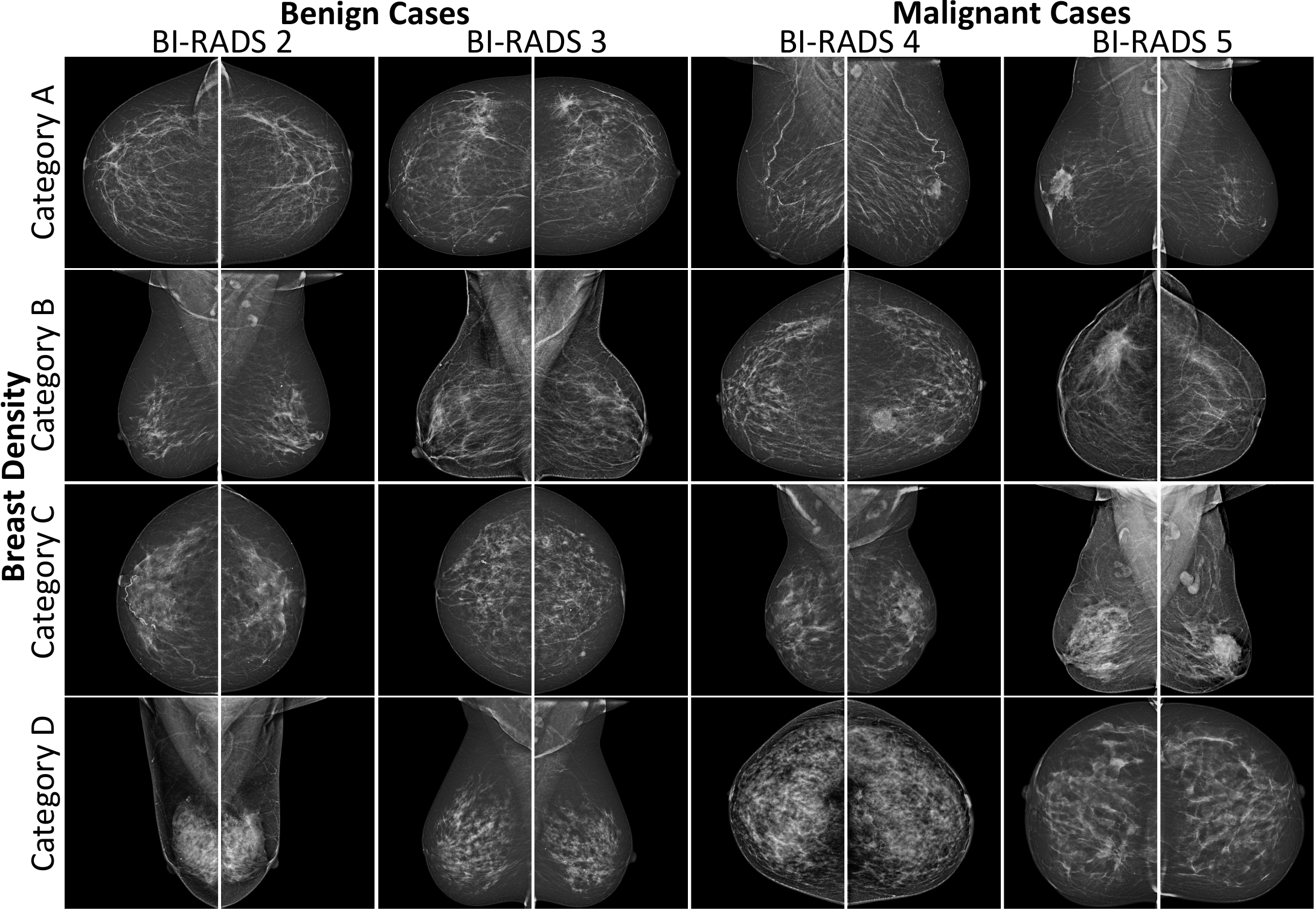}\vspace{-5pt}
    \caption{\textbf{Representative benign and malignant mammograms.}}\vspace{-10pt}
    \label{fig:samples}
\end{figure}

Our contribution can be summarized as follows:
\begin{itemize}
\item \textbf{LUMINA dataset.} Unlike prior resources that are single‑vendor or film‑based (e.g., CBIS‑DDSM, INbreast), LUMINA is explicitly multi‑vendor and energy‑annotated FFDM at 12–14‑bit depth, with pathology‑confirmed outcomes, per‑breast BI‑RADS, and density labels. This combination enables controlled stress‑tests for vendor/energy shifts and evaluation of single‑ vs. two‑view modeling at clinically used resolutions. Fig.~\ref{fig:samples} presents representative benign and malignant mammograms.
\item \textbf{Foreground-only histogram harmonization.} A simple, vendor-agnostic histogram matching that excludes background pixels and preserves lesion contrast. 
\item \textbf{Three-task benchmark with multi-view modeling.} Unified evaluation on (i) pathology, (ii) BI-RADS risk grouping (binary and 3-class), and (iii) density; two-view models outperformed single-view for diagnosis, with EfficientNet-B0 reaching 93.61\% AUC; EfficientNet-B0 also obtains the highest AUC (93.97\%) for BI-RADS, and Swin-T yields the best AUC (89.10\%) for density. 

\item \textbf{Harmonization improves models and attention.} Foreground histogram matching consistently raises AUC/ACC across settings (Fig.~\ref{fig:ablation} in Sec.~\ref{sec:Ablation Study And Discussion}) and yields more focal Grad-CAMs around suspicious regions (Fig.~\ref{fig:gradcam} in Sec.~\ref{sec:Ablation Study And Discussion}), indicating better localization of lesion-related signal. 
\end{itemize}

\section{Related Works}
\begin{table*}[tb]
\centering
\caption{\textbf{Comparison of publicly available mammography datasets.}}\vspace{-5pt}
\label{tab:dataset_comparison}
\begin{tabular}{lccccccc}
\toprule
&  &  & & \textbf{Pathology}&\textbf{BI-RADS}&\textbf{Breast}&\textbf{Acquisition}\\ 
\multirow{-2}{*}{\textbf{Dataset}} & \multirow{-2}{*}{\textbf{Origin}} & \multirow{-2}{*}{\textbf{Patients}} & \multirow{-2}{*}{\textbf{Images}}& \textbf{Labels}&\textbf{Assessment}&\textbf{Density}& \textbf{Technique}\\ 
\midrule
MIAS~\cite{suckling1994mammographic} & UK & 161 &322&Yes&No&No&SFM\\
DDSM~\cite{bowyer1996digital} & USA & 2,620&10,480 & Yes&Yes&Yes&SFM \\
CBIS-DDSM~\cite{lee2017curated} & USA & 2,620&10,239 & Yes&Yes&Yes&SFM \\
INbreast~\cite{moreira2012inbreast} & Portugal & 115&410&No&Yes&Yes&FFDM\\
VinDR-Mammo~\cite{pham2022vindr} & Vietnam & 5,000 &20,000 & No&Yes&Yes&FFDM \\
RSNA~\cite{carr2022rsna} & USA & 1,970 & 9,594 & Yes &  Limited* & Yes &FFDM \\
CMMD~\cite{cai2023online} & China & 1,775  & 3,728 & Yes &No &No & FFDM \\
CDD-CESM~\cite{khaled2022categorized} & Egypt & 326 & 1,003 & Yes&Yes&Yes&CESM\\
KAU-BCMD~\cite{alsolami2021king} & Saudi Arabia & 442&1,774 & No & Yes & No &FFDM\\
\textbf{LUMINA (Ours)} & \textbf{Turkey} & \textbf{468} &\textbf{1,824}& \textbf{Yes}& \textbf{Yes} & \textbf{Yes} & \textbf{FFDM}\\
\bottomrule
\multicolumn{8}{l}{\footnotesize{*RSNA provides simplified BI-RADS labels (0: follow-up, 1: negative, 2: normal) instead of the full BI-RADS scores from 0 to 6.}}\\
\end{tabular}\vspace{-10pt}
\end{table*}

\noindent\textbf{Mammography datasets:} 
The Mammographic Image Analysis Society (MIAS) database~\cite{suckling1994mammographic}, the Digital Database for Screening Mammography (DDSM)~\cite{bowyer1996digital}, and its curated extension CBIS-DDSM~\cite{lee2017curated} have been widely used in early computer-aided detection system development, providing digitized film mammograms with associated lesion annotations and pathology verification. However, these datasets contain relatively low-resolution screen-film mammography (SFM) scans, limiting their applicability to current clinical practice. To address these limitations, several institutions have released modern full-field digital mammography (FFDM) datasets such as INbreast~\cite{moreira2012inbreast}, VinDR-Mammo~\cite{pham2022vindr}, RSNA Screening Mammography~\cite{carr2022rsna}, Chinese Mammography Database (CMMD)~\cite{cai2023online}, CDD-CESM~\cite{cai2023online}, KAU-BCMD~\cite{alsolami2021king}. 
Despite recent progress, many existing mammography datasets face limitations, including small sample sizes, reliance on film-based scans, or incomplete integration of radiological and pathological labels. Compared with these resources in Table~\ref{tab:dataset_comparison}, LUMINA provides high-resolution multi-view FFDM images with pathology-confirmed outcomes, expert-provided BI-RADS risk categories, and full breast density annotations.  These allow LUMINA to fill critical gaps in existing datasets and establish a valuable benchmark for developing and evaluating AI algorithms in breast cancer imaging.

\noindent\textbf{Mammography classification:} 
CNN-based studies adapted architectures such as AlexNet~\cite{omonigho2020breast}, VGG~\cite{montaha2021breastnet18,rathinam2024adaptive}, and ResNet~\cite{chen2018fine,priya2021resnet} for mammography classification, showing clear improvements over handcrafted features~\cite{dhungel2015deep,arevalo2016representation}. More specialized designs, such as multi-view CNNs that jointly model cranio-caudal (CC) and mediolateral oblique (MLO) views, have further enhanced diagnostic performance~\cite{shen2019deep,lotter2021robust}. Recently, transformer-based architectures have been explored, leveraging self-attention mechanisms to capture long-range dependencies across mammographic views and regions~\cite{ayana2023vision}. 

\noindent\textbf{Medical image harmonization:} Harmonizing medical images across different scanners, acquisition protocols, and vendors is critical to enabling robust and generalizable AI systems in healthcare. Statistical approaches such as ComBat~\cite{orlhac2022guide} employ an empirical Bayes framework to correct batch-effects for MRI images by adjusting location and scale parameters of extracted imaging features, and have been shown to significantly reduce inter-site variability while preserving biological signal.  Deep learning approaches, such as HarmoFL~\cite{jiang2022harmofl}, further address feature drift by normalizing frequency domain amplitudes and guiding model convergence in federated learning setups across heterogeneous clients. 

 \noindent\textbf{How we differ:} Prior mammography studies typically address \emph{one} task (diagnosis or BI-RADS or density) within a constrained imaging domain, often single‑vendor FFDM or film‑based datasets. In contrast, \textbf{LUMINA} couples (i) \emph{multi‑vendor} FFDM with energy metadata, (ii) \emph{unified} evaluation across \emph{three} clinically relevant tasks, and (iii) a \emph{foreground‑only, pixel‑space} CDF alignment that reduces vendor/energy intensity drift while preserving lesion morphology. Unlike ComBat‑style feature harmonizers or federated normalization methods, our approach is \emph{model‑agnostic}, runs as a lightweight pre‑processing step, and consistently improves AUC and attention localization across backbones (Fig.~\ref{fig:ablation} in Sec.~\ref{sec:Ablation Study And Discussion}). We further report view‑ablations showing that \emph{two‑view} models outperform four‑view configurations while maintaining parameter efficiency (Table~\ref{tab:pathology} in Sec.~\ref{sec:Benchmark Results}).

\section{LUMINA Dataset Introduction}~\label{sec:LUMINA Dataset Introduction}
\begin{figure}[tb]
    \centering
    \subfloat[Age distribution.\label{fig:age}]{
    \includegraphics[width=0.33\linewidth]{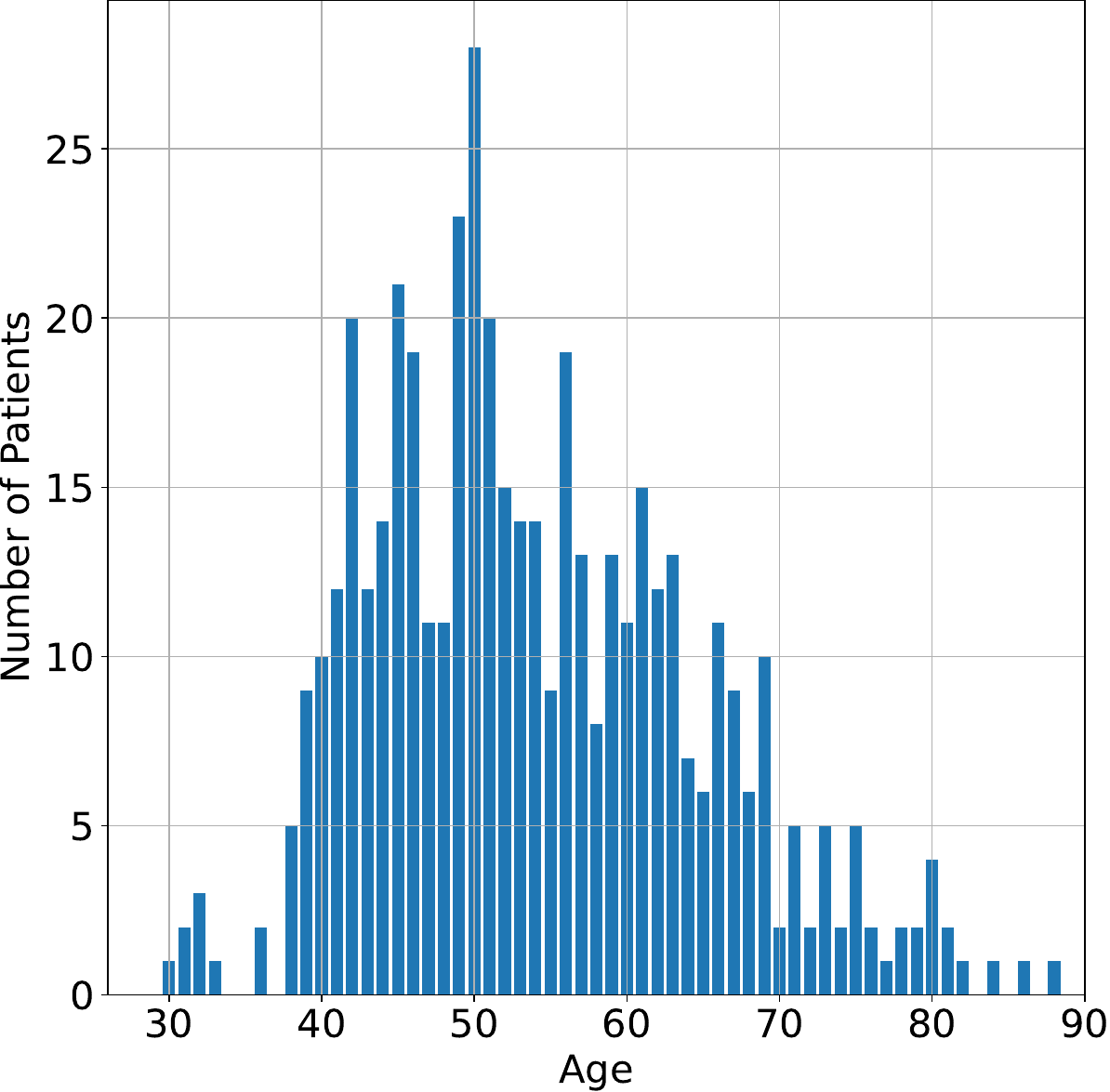}
    }
    \subfloat[BI-RADs distribution.\label{fig:birads distribution}]{
    \includegraphics[width=0.33\linewidth]{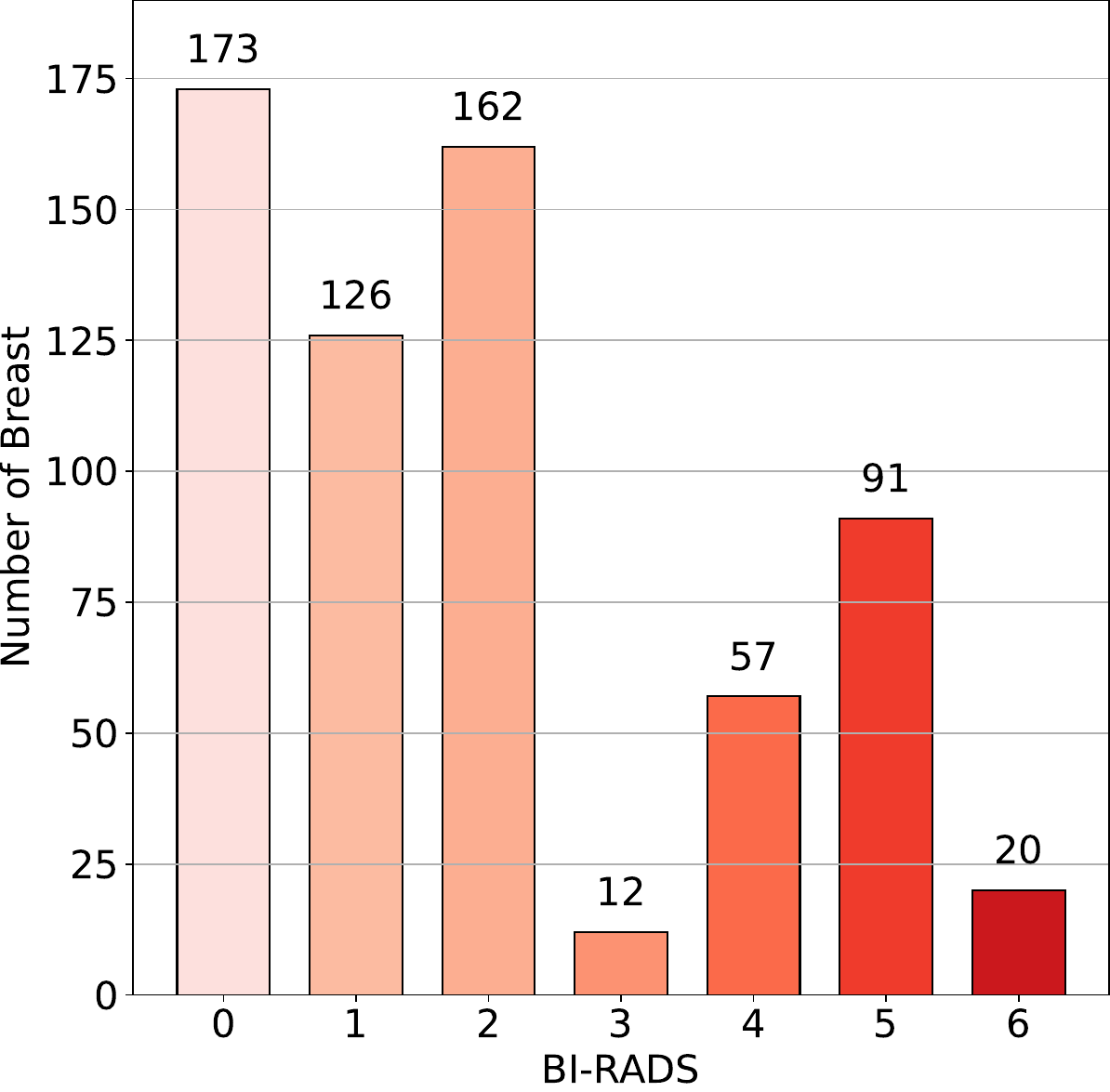}
    }
    \subfloat[Density distribution.\label{fig:density distribution}]{
    \includegraphics[width=0.33\linewidth]{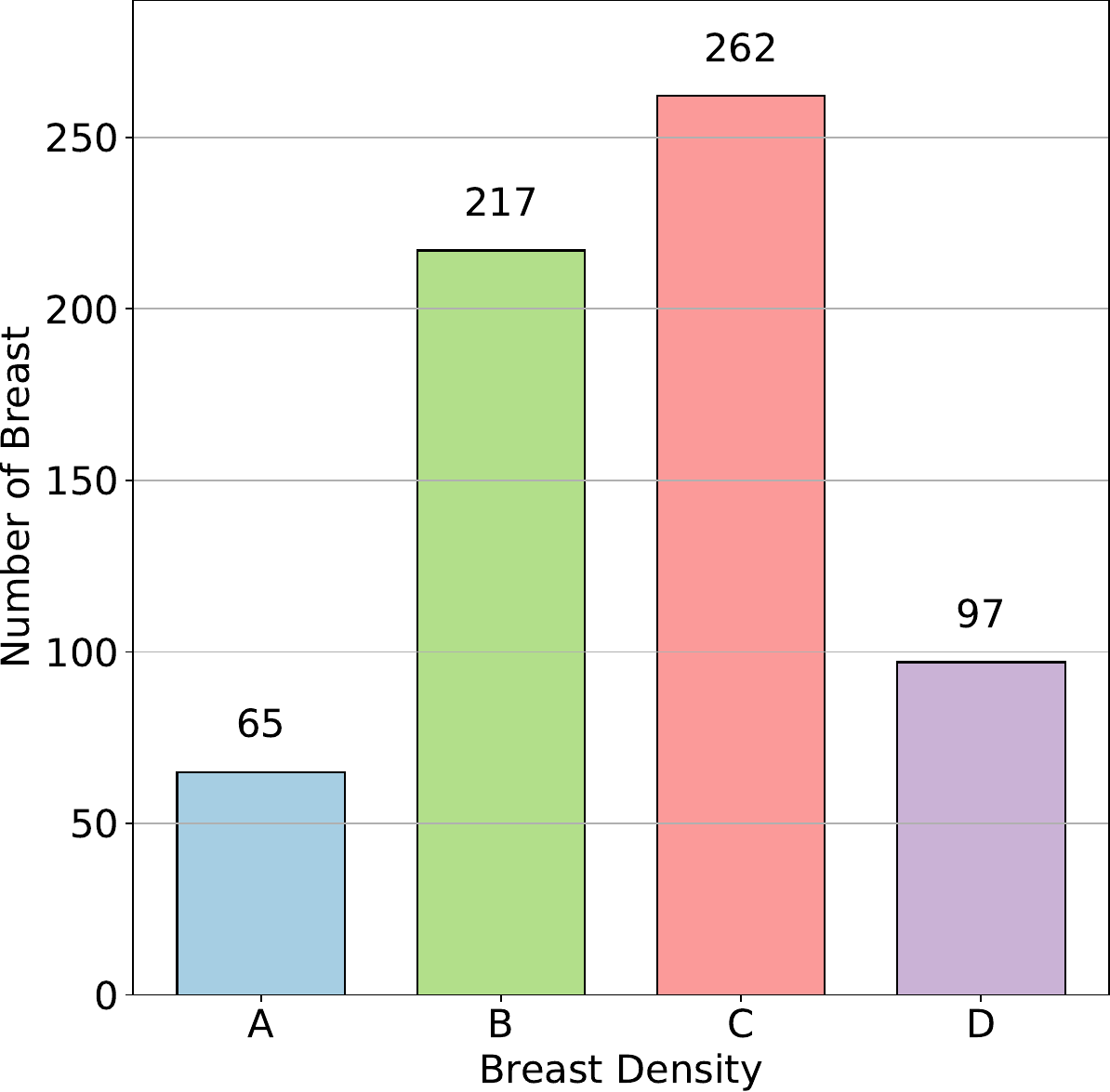}
    }\vspace{-5pt}
    \caption{\textbf{Age, BI-RADS, and breast distribution.}}\vspace{-15pt}
    \label{fig:distribution}
\end{figure}

The LUMINA dataset is a curated collection of FFDM developed in collaboration with the University of Health Sciences, Prof. Dr. Cemil Tascioglu City Hospital, Istanbul, Turkey. 
It consists of 1,824 images from 468 patients, including 960 images from 250 benign patients and 864 images from 218 malignant patients, with malignancy confirmed by pathology. Patient ages range from 30 to 88 years, as shown in Fig.~\ref{fig:age}. Most cases include four-view mammograms (CC and MLO views for both breasts), while a subset contains only two views due to incomplete acquisition or prior surgical history. Images are accompanied by expert-provided annotations, including BI-RADS risk assessments (categories 0–6) and breast density classifications (A–D), alongside the pathology-confirmed outcome, as summarized in Fig.~\ref{fig:distribution}. While the dataset contains 1,824 images in total, only 1,282 images have fully interpretable BI-RADS and density annotations. This discrepancy arises because malignancy typically appears unilaterally. Only the breast exhibiting malignancy (as confirmed by pathology) is included in the malignant class, while the contralateral breast is discarded.

The mammography images were collected from six mammography imaging systems: IMS, Metaltronica, FUJIFILM Corporation, Siemens, Carestream Health, and GE Medical Systems. Details of these vendors are summarized in Table~\ref{tab:vendor}. Mammograms are stored in DICOM format with native resolutions ranging from $2364\times 2964$ to $4800 \times 6000$ pixels and a bit depth of 12–14 bits per pixel, preserving fine diagnostic details. Most images use the MONOCHROME2 format (higher pixel values correspond to brighter regions), while FUJIFILM systems use MONOCHROME1, where the intensity mapping is inverted. These MONOCHROME1 images were converted to MONOCHROME2 to ensure consistent visualization across the dataset.

\begin{table}[tb]
    \caption{\textbf{Vendor distribution.}}\vspace{-5pt}
    \centering
    \begin{tabular}{l|cccc}
    \toprule
         Vendor&\# Patients&\# Images&Energy\\
         \midrule
         IMS&341&1326&High\\
         Metaltronica&91&354&Low\\
         FUJIFILM Corporation&29&116&Low\\
         SIEMENS&4&16&Low\\
         Carestream Health&1&4&Low\\
         GE MEDICAL SYSTEMS&2&8&High\\
         \midrule
        Total&468&1824\\
         \bottomrule
    \end{tabular}\vspace{-10pt}
    \label{tab:vendor}
\end{table}

Because mammograms have significant domain shifts as mentioned before, we applied a foreground‑only histogram harmonization (Section~\ref{sec:Mammogram Image Enhancement via Histogram Matching}) to align intensity distributions across vendors toward a unified low-energy style while preserving lesion integrity.  Representative samples from LUMINA are shown in Fig.~\ref{fig:samples}.

\begin{figure*}[tb]
\centerline{\includegraphics[width=.75\linewidth]{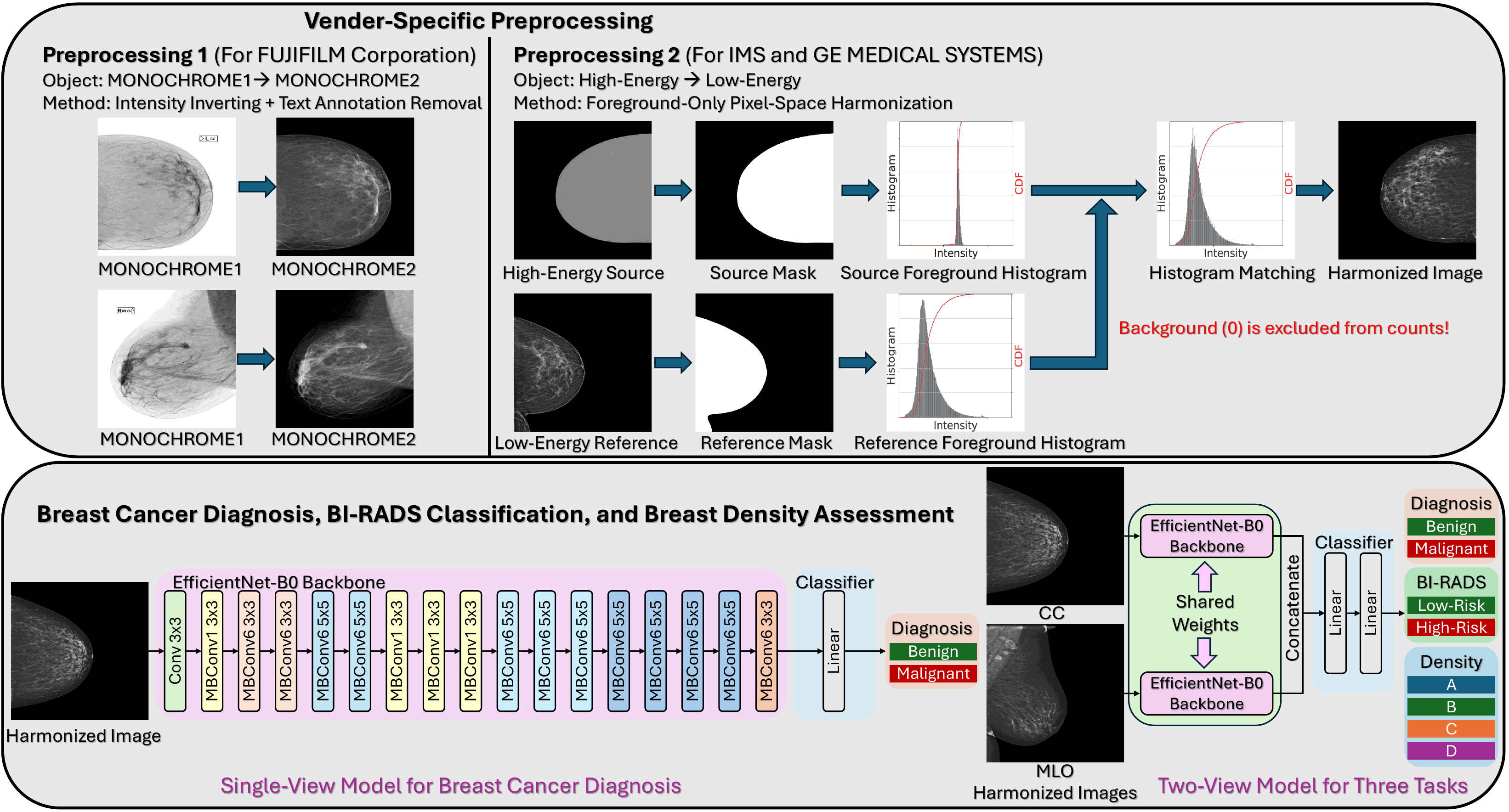}}\vspace{-5pt}
\caption{\textbf{LUMINA pipeline.} The EfficientNet-B0 backbone can be replaced by other backbones (ResNet-50, DenseNet-121, and Swin-T). Two‑view \emph{shared‑backbone} reached accuracy comparable to independent‑backbone with 48\% less parameters (Table~\ref{tab:pathology ablation}) and outperformed four‑view variants (Table~\ref{tab:pathology}).}\vspace{-15pt}
\label{fig:pipeline}
\end{figure*}

\begin{table}[tb]
    \caption{\textbf{Breast cancer diagnosis task data distribution.} Numbers outside and inside brackets are cases and images, respectively.}\vspace{-5pt}
    \centering
    \begin{tabular}{l|ccc}
    \toprule
         View&Benign&Malignant&Total\\
         \midrule
         Single&592 (592)&344 (344)&936 (936)\\
         Two&296 (592)&172 (344)&468 (936)\\
         \bottomrule
    \end{tabular}\vspace{-10pt}
    \label{tab:pathology distribution}
\end{table}

\subsection{Evaluation Protocol}
We evaluate the LUMINA dataset using three clinically relevant classification tasks: breast cancer diagnosis, BI-RADS risk assessment, and breast density prediction. All tasks are performed on standard two-view mammograms (CC and MLO) unless otherwise noted.

\noindent \textbf{Breast cancer diagnosis:} 
This is the primary task in this study, as it is clinically important for early detection and risk assessment. Models are trained using labels confirmed by pathology to distinguish benign from malignant cases. Images with BI-RADS 0 are excluded, as BI-RADS 0 indicates incomplete assessments requiring additional imaging. We evaluate single-view and two-view configurations. Only images corresponding to malignant findings are included in the single- and two-view malignancy subsets, and other images from the malignant patients are discarded. As Table~\ref{tab:pathology distribution} presents, 936 images are used for single-view evaluation and 468 cases (936 images) for two-view evaluation.

\noindent \textbf{BI-RADS classification:} 
This task focuses on predicting BI-RADS risk categories to mimic radiologists’ assessments. Only the standard two-view images are used, as BI-RADS annotations are assigned per breast without pathological confirmation. We evaluate two schemes: (1) binary classification, distinguishing low-risk (BI-RADS 1–3) from high-risk (BI-RADS 4–6), and (2) three-class classification, grouping cases into low-risk (BI-RADS 1–2), intermediate-risk (BI-RADS 3–4), and high-risk (BI-RADS 5–6). Images with BI-RADS 0 are excluded as in the previous task. As a result, 468 cases (936 images) are used for evaluation.

\noindent \textbf{Breast density assessment:} 
Accurate breast density prediction is essential because higher-density breasts (C–D) can obscure lesions and reduce mammography sensitivity. Models are trained to predict the four density categories (A–D) using two-view images for each breast. Since BI-RADS 0 does not affect density determination, these images are included. The evaluation includes 641 cases (1,282 images), enabling development of models that reliably predict breast density across multiple vendors and support risk stratification and AI-assisted interpretation.

\section{Pixel-Space Harmonization}\label{sec:Mammogram Image Enhancement via Histogram Matching}

Modern FFDM systems typically acquire images using either low-energy or high-energy X-ray settings. Low-energy mammograms are optimized for soft-tissue contrast, enhancing the visibility of subtle lesions, while high-energy mammograms emphasize denser structures such as calcification. However, these acquisition differences, combined with vendor-specific processing, introduce significant intensity and contrast variations across devices. Such inconsistencies can degrade the generalization performance of deep learning models, as models trained on one vendor’s imaging style may perform poorly on another. Therefore, harmonizing intensity distributions across systems is critical for developing vendor-agnostic and clinically reliable AI models.

\begin{figure}[tb]
   \centering
    \includegraphics[width=.8\linewidth]{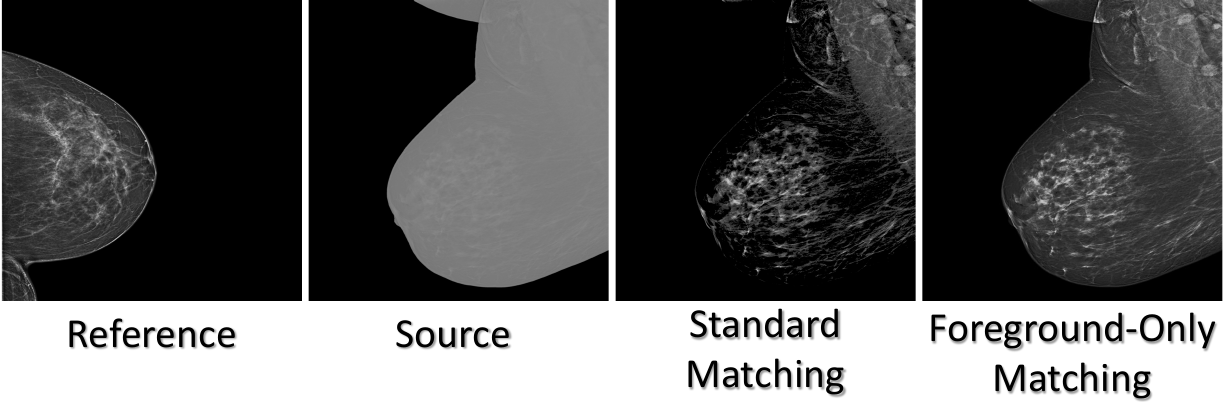}\vspace{-5pt}
\caption{\textbf{Background influence.} Standard histogram matching is degraded by large black background regions, whereas foreground-only histogram matching remains unaffected.}
\vspace{-15pt}
\label{fig:degrade}
\end{figure}

To address these discrepancies, the LUMINA dataset was harmonized using a histogram-matching-based approach. This method aligns the intensity distribution of each image to a reference histogram, thereby improving consistency across multi-vendor datasets while preserving fine diagnostic details. Histogram matching~\cite{rolland2000fast,shen2007image,tu2013histogram,shapira2013multiple} transforms the intensity distribution of a source image to match that of a reference image. However, unlike nature images, mammograms contain large black background regions. Including these areas in the histogram can degrade the quality of the matching, as Fig.~\ref{fig:degrade} shows. Therefore, these regions are excluded before applying histogram matching. Let \(M_s = \{(x,y)\mid \mathbf{I}_s(x,y) > 0\}\) and \(M_r = \{(x,y)\mid \mathbf{I}_r(x,y) > 0\}\) denote the sets of foreground (breast) pixels in the source image $\mathbf{I}_s$ and reference image $\mathbf{I}_r$, respectively. We compute histograms only over these foreground pixels to avoid bias from black background regions:
\begin{equation}
    H_s(k) = \#\{(x,y)\in M_s \mid \mathbf{I}_s(x,y) = k\}, \end{equation}
    \begin{equation}
    H_r(k) = \#\{(x,y)\in M_r \mid \mathbf{I}_r(x,y) = k\},
\end{equation}
and the total foreground pixel counts:

\begin{equation}
    N_s = \sum_{k} H_s(k) = |M_s|,
\end{equation}    
\begin{equation}
    N_r = \sum_{k} H_r(k) = |M_r|.
\end{equation}
We then form the normalized cumulative distribution functions (CDFs) over the foreground intensities:
\begin{equation}
    C_s(p) = \sum_{k=1}^{p} H_s(k), \quad \bar{C}_s(p) = \frac{C_s(p)}{N_s},\label{eq:cdfsource}
\end{equation}
\begin{equation}
    C_r(q) = \sum_{k=1}^{q} H_r(k), \quad \bar{C}_r(q) = \frac{C_r(q)}{N_r},\label{eq:cdfreference}
\end{equation}
where the summation starts from intensity 1, as intensity 0 represents the background and is excluded.
The mapping function \(\mathcal{T}(\cdot)\) is defined by matching these CDFs:
\begin{equation}
    \mathcal{T}(p) = {\arg\min}_{q\in\{1,\dots,2^b{-}1\}} \, \big|\bar{C}_s(p) - \bar{C}_r(q)\big|.\label{eq:mapping}
\end{equation}
Finally, the harmonized image \(\mathbf{I}_o\) is obtained by applying \(\mathcal{T}\) to foreground pixels and leaving background pixels as zero:
\begin{equation}
    \mathbf{I}_o(x,y) =
    \begin{cases}
      \mathcal{T}\!\big(\mathbf{I}_s(x,y)\big), & (x,y)\in M_s,\\
      0, & \text{otherwise.}
    \end{cases}
    \label{eq:histogram_mapping_masked}
\end{equation}
Fig.~\ref{fig:pipeline} illustrates an example: the source image exhibits high contrast, and its CDF is adjusted to follow that of the reference image, resulting in improved intensity consistency.

\noindent\textbf{Practical notes:} We derive the reference histogram from a representative subset of low‑energy FFDM images to stabilize soft‑tissue contrast. Foreground masking is computed by thresholding at intensity $>0$ after MONOCHROME1$\to$2 conversion to avoid background bias (Fig.~\ref{fig:pipeline}). Histogram computation uses 12‑bit bins to preserve detail. This pre‑processing is model‑agnostic and applied identically across tasks and backbones. The harmonization algorithm is summarized in Algorithm~\ref{alg:masked-cdf}.

\begin{algorithm}[t]
\caption{Foreground-only CDF Matching for Vendor/Energy Harmonization}
\label{alg:masked-cdf}
\small
\begin{algorithmic}[1]
\REQUIRE Source image $\mathbf{I}_s$, reference image $\mathbf{I}_r$ (low-energy style), bit depth $b$
\STATE Convert MONOCHROME1 to MONOCHROME2 if needed; remove textual burn-ins (see Fig.~\ref{fig:pipeline}). 
\STATE Foreground masks: $M_s \leftarrow \{(x,y)\!\mid\!\mathbf{I}_s(x,y)>0\}$, $M_r \leftarrow \{(x,y)\!\mid\!\mathbf{I}_r(x,y)>0\}$
\STATE Compute histograms over foreground: $H_s, H_r \in \mathbb{N}^{2^b}$; set $H_s(0)=H_r(0)=0$
\STATE Compute normalized CDFs $\bar C_s,\bar C_r$ (Eqs.~(\ref{eq:cdfsource})--(\ref{eq:cdfreference}))
\FOR{$p=1$ \TO $2^b{-}1$}
  \STATE $\mathcal{T}(p)\leftarrow \arg\min_{q\in\{1,\dots,2^b{-}1\}} |\bar C_s(p) - \bar C_r(q)|$ \hfill \COMMENT{monotone mapping}
\ENDFOR
\STATE Initialize $\mathbf{I}_o \leftarrow \mathbf{0}$; \textbf{for} $(x,y)\in M_s$: $\mathbf{I}_o(x,y)\leftarrow \mathcal{T}(\mathbf{I}_s(x,y))$
\STATE \textbf{return} $\mathbf{I}_o$
\end{algorithmic}
\end{algorithm}

\section{Deep Learning Approaches for Multi-View Mammography Classification}
We adopt EfficientNet-B0~\cite{tan2019efficientnet} as the representative backbone, though alternative architectures such as ResNet-50~\cite{he2016deep}, DenseNet-121~\cite{huang2017densely}, and Swin-T~\cite{liu2021swin} can also be utilized. All backbones are pretrained on the ImageNet-1K dataset~\cite{deng2009imagenet} to provide robust initialization. In the single-view setting, the model is fine-tuned through conventional transfer learning. In the two-view setting, the CC and MLO images of the same breast are independently processed through a shared-weight backbone. The resulting feature embeddings are concatenated and passed through a series of fully connected layers for classification. This shared-weight configuration enables efficient parameter utilization and promotes consistent feature learning across views while leveraging complementary information from both projections for improved diagnostic accuracy. The BI-RADS classification and breast density prediction models follow the same two-view architecture, differing only in the output layer of the classification head.

\begin{table*}[tb]
    \caption{\textbf{Mammogram cancer diagnosis benchmark.} The best and second-best results are highlighted in green and yellow, respectively.}\vspace{-5pt}
    \centering
    \setlength{\tabcolsep}{1pt}
    \begin{tabular}{l|cccccccccc}
    \toprule         \textbf{View}&\textbf{Model}&\textbf{Input}&\textbf{Params}&\textbf{Flops}&\textbf{ACC(\%)}&\textbf{AUC(\%)}&\textbf{Precision(\%)}&\textbf{Recall(\%)}&\textbf{F1(\%)}&\textbf{Specificity(\%)}\\
         \midrule
         \multirow{8}{*}{Single}&ResNet-50 & $224^2$ & 23.51M & 4.13G & 79.49$\pm$5.99 & 87.42$\pm$4.21 & \best{81.29$\pm$10.90} & 61.97$\pm$18.17 & 67.52$\pm$13.09 & \best{89.68$\pm$9.63} \\
         &DenseNet-121 & $224^2$ & 6.95M & 2.90G & 79.49$\pm$3.37 & 87.14$\pm$3.08 & \secondbest{80.45$\pm$9.45} & 62.55$\pm$17.56 & 67.79$\pm$9.57 & \secondbest{89.38$\pm$7.52} \\
         &EfficientNet-B0 & $224^2$ & 4.01M & 0.41G & \best{83.12$\pm$4.13} & \best{90.66$\pm$4.21} & 79.78$\pm$6.31 & \best{73.89$\pm$15.77} & \best{75.50$\pm$7.95} & 88.51$\pm$5.00 \\
         &Swin-T & $224^2$ & 27.52M & 3.13G & \secondbest{81.84$\pm$5.92} & \secondbest{88.62$\pm$5.18} & 78.27$\pm$8.63 & \secondbest{71.83$\pm$12.10} & \secondbest{74.16$\pm$8.36} & 87.66$\pm$7.45 \\\cline{2-11}
         
         &ResNet-50 & $512^2$ & 23.51M & 21.58G & 77.68$\pm$7.18 & 86.69$\pm$4.37 & 82.79$\pm$14.09 & 59.36$\pm$28.73 & 61.41$\pm$22.78 & 88.38$\pm$12.29\\
         &DenseNet-121 & $512^2$ & 6.95M & 15.14G & 79.38$\pm$5.55 & 90.52$\pm$3.61 & 81.32$\pm$16.39 & 68.09$\pm$21.96 & 69.22$\pm$11.98 & 85.99$\pm$14.42\\
         &EfficientNet-B0 & $512^2$ & 4.01M & 2.13G & \best{86.43$\pm$3.85} & \best{92.13$\pm$4.21} & \best{86.61$\pm$5.58} & \best{75.03$\pm$9.46} & \best{80.00$\pm$6.35} & \best{93.08$\pm$3.31} \\
         &Swin-T & $512^2$ & 27.52M & 16.47G & \secondbest{84.94$\pm$3.57} & \secondbest{91.34$\pm$4.22} & \secondbest{85.33$\pm$4.48} & \secondbest{71.55$\pm$10.46} & \secondbest{77.38$\pm$6.35} & \secondbest{92.73$\pm$2.55}\\\midrule  
         
          \multirow{8}{*}{Two}&ResNet-50&$224^2$&24.03M&8.26G& 76.70$\pm$7.80 & 88.71$\pm$2.82 & \secondbest{87.99$\pm$14.28} & 48.87$\pm$27.42 & 55.53$\pm$24.63 & 92.89$\pm$9.12\\
          &DenseNet-121&$224^2$& 7.22M&5.80G& \secondbest{84.40$\pm$1.64} & 89.86$\pm$1.21 & \best{88.67$\pm$5.88} & \secondbest{66.81$\pm$7.49} & \secondbest{75.68$\pm$3.62} & \best{94.59$\pm$3.63} \\
          &EfficientNet-B0&$224^2$&4.34M&0.82G & \best{86.11$\pm$2.47} & \best{92.99$\pm$3.04} & 86.83$\pm$2.93 & \best{73.26$\pm$5.63} & \best{79.39$\pm$4.18} & \secondbest{93.58$\pm$1.26} \\
          &Swin-T&$224^2$&27.72M&6.26G& 81.22$\pm$9.77 & \secondbest{91.99$\pm$3.40} & 63.41$\pm$32.06 & 66.57$\pm$33.56 & 64.92$\pm$32.73 & 89.88$\pm$5.93\\\cline{2-11}
          
          &ResNet-50&$512^2$&24.03M&43.16G& 72.62$\pm$14.72 & 87.56$\pm$5.14 & 70.66$\pm$16.28 & 71.45$\pm$27.02 & 64.29$\pm$17.57 & 73.23$\pm$30.35\\
          &DenseNet-121&$512^2$& 7.22M&30.29G& 72.46$\pm$11.81 & 88.33$\pm$5.46 & 64.93$\pm$13.99 & \best{79.01$\pm$13.83} & 68.88$\pm$5.96 & 68.56$\pm$25.60\\
          &EfficientNet-B0&$512^2$ &4.34M&4.27G& \best{85.04$\pm$5.92} & \best{93.54$\pm$3.88} & \best{91.34$\pm$11.37} & 69.87$\pm$22.03 & \secondbest{75.75$\pm$12.85} & \best{93.91$\pm$9.75} \\
          &Swin-T&$512^2$&27.72M&32.94G & \secondbest{83.96$\pm$4.25} & \secondbest{92.42$\pm$4.27} & \secondbest{80.92$\pm$10.00} & \secondbest{76.15$\pm$9.13} & \best{77.67$\pm$5.36} & \secondbest{88.50$\pm$6.91}\\
          
         \bottomrule
    \end{tabular}\vspace{-10pt}
    \label{tab:pathology}
\end{table*}

\begin{table*}[tb]
    \caption{\textbf{BI-RADS classification benchmark.}
    }\vspace{-5pt}
    \centering
    \begin{tabular}{lc|ccc|ccc}
    \toprule
         &&\multicolumn{3}{c|}{\textbf{Two-Class (BI-RADS 1/2/3 VS 4/5/6)}}&\multicolumn{3}{c}{\textbf{Three-Class (BI-RADS 1/2 VS 3/4 VS 5/6)}}\\
         \multirow{-2}{*}{\textbf{Model}}&\multirow{-2}{*}{\textbf{Input}}&\textbf{ACC(\%)}&\textbf{AUC(\%)}&\textbf{F1(\%)}&\textbf{ACC(\%)}&\textbf{AUC(\%)}&\textbf{F1(\%)}\\
         \midrule
         ResNet-50&$224^2$& 71.98$\pm$12.34 & 88.23$\pm$3.52 & \secondbest{62.45$\pm$10.33}& 69.02$\pm$5.14 & 79.58$\pm$3.00 & \best{55.36$\pm$1.97}\\
          DenseNet-121&$224^2$ & 77.55$\pm$3.67 & 87.89$\pm$3.52 & 58.98$\pm$14.70& 68.15$\pm$3.12 & 80.90$\pm$3.74 & 48.63$\pm$3.49\\
          EfficientNet-B0&$224^2$& \best{83.56$\pm$4.76} & \best{92.80$\pm$4.13} & \best{74.18$\pm$9.32}& \best{71.78$\pm$5.03} & \best{83.27$\pm$4.91} & \secondbest{55.15$\pm$4.65} \\
          Swin-T&$224^2$& \secondbest{79.29$\pm$9.01} & \secondbest{90.80$\pm$4.61} & 60.82$\pm$31.02& \secondbest{70.51$\pm$4.45} & \secondbest{81.69$\pm$5.55} & 49.55$\pm$6.85\\\midrule
          ResNet-50&$512^2$& 75.22$\pm$4.46 & 89.08$\pm$5.42 & 64.78$\pm$9.96& 66.66$\pm$3.10 & 78.92$\pm$4.91 & 50.71$\pm$3.73\\
          DenseNet-121&$512^2$& 74.56$\pm$10.73 & 88.76$\pm$3.95 & 59.09$\pm$24.68 & 64.97$\pm$5.95 & 80.54$\pm$4.29 & \secondbest{52.50$\pm$3.48}\\
          EfficientNet-B0&$512^2$& \best{85.48$\pm$3.04} & \best{92.75$\pm$4.10} & \best{76.83$\pm$5.45}& \best{71.13$\pm$5.22} & \best{82.85$\pm$4.41} & \best{55.73$\pm$3.22}\\
          Swin-T&$512^2$& \secondbest{83.31$\pm$6.21} & \secondbest{91.44$\pm$4.41} & \secondbest{76.29$\pm$7.12}& \secondbest{70.07$\pm$5.02} & \secondbest{81.68$\pm$5.88} & 47.98$\pm$5.66\\

         \bottomrule
    \end{tabular}\vspace{-10pt}
    \label{tab:birads}
\end{table*}

\begin{table}[tb]
    \caption{\textbf{Density classification benchmark.} 
    }\vspace{-5pt}
    \centering
    \setlength{\tabcolsep}{1pt}
    \begin{tabular}{lcccc}
    \toprule
         \textbf{Model}&\textbf{Input}&\textbf{ACC(\%)}&\textbf{AUC(\%)}&\textbf{F1(\%)}\\
         \midrule
         ResNet-50&$224^2$& 64.74$\pm$3.33 & 85.66$\pm$1.62 & \secondbest{56.94$\pm$6.43}  \\
         DenseNet-121&$224^2$ & \secondbest{66.29$\pm$4.20} & \secondbest{86.39$\pm$2.92} & 56.70$\pm$2.97  \\
         EfficientNet-B0&$224^2$& 59.43$\pm$5.23 & 84.72$\pm$3.54 & 48.25$\pm$5.68 \\
         Swin-T&$224^2$& \best{70.35$\pm$4.60} & \best{89.43$\pm$2.22} & \best{65.28$\pm$6.48} \\\midrule
         ResNet-50&$512^2$& 65.37$\pm$4.07 & \secondbest{87.06$\pm$2.10} & 58.29$\pm$7.57 \\
         DenseNet-121&$512^2$& 66.31$\pm$5.31 & 86.60$\pm$3.76 & 61.42$\pm$7.62\\
         EfficientNet-B0&$512^2$& \secondbest{67.54$\pm$3.89} & 86.83$\pm$2.80 & \best{63.10$\pm$4.90}\\
         Swin-T&$512^2$& \best{69.41$\pm$7.23} & \best{89.14$\pm$3.36} & \secondbest{61.50$\pm$12.60}\\

         \bottomrule
    \end{tabular}\vspace{-10pt}
    \label{tab: density}
\end{table}

\section{Experimental Results}
\subsection{Experimental Setup}\label{sec:Experimental Setup}

\noindent\textbf{Baseline fairness:} All backbones used the same data splits, augmentations, and training schedule to ensure comparability. We applied horizontal flip and resizing, replicate grayscale to three channels, with identical optimization settings across tasks. CNN backbones (ResNet-50, DenseNet-121, EfficientNet-B0) were trained by AdamW optimizer~\cite{loshchilov2017decoupled} with an initial learning rate of $1\times10^{-3}$, while Swin-T started with $1\times10^{-5}$. All models were trained for 100 epochs with a weight decay of $1\times10^{-5}$, the learning rate decayed by a factor of $0.1$ every 30 epochs, and the selection of the model by the best validation AUC under 5-fold cross validation. This protocol avoids method-specific tricks and isolates architectural effects. PyTorch and CUDA determinism flags are enabled for all reported runs. All experiments were implemented in PyTorch on a server with 8 NVIDIA A6000 GPUs. The LUMINA dataset and source code will be released upon paper acceptance.

\noindent\textbf{Evaluation:} We reported accuracy (ACC), area under the ROC curve (AUC), F1-score, sensitivity (recall), precision, and specificity for breast cancer diagnosis, and ACC, AUC, and F1 for the three-class BI-RADS and density classification (macro-AUC and macro-F1 for multi-class). AUC is considered the primary metric from a clinical perspective. Results were reported as mean$\pm$std over 5 folds.

\subsection{Benchmark Results}\label{sec:Benchmark Results}
\begin{figure}[tb]
   \centering
    \subfloat[Single-view features, $224^2$.]{
    \includegraphics[width=0.48\linewidth]{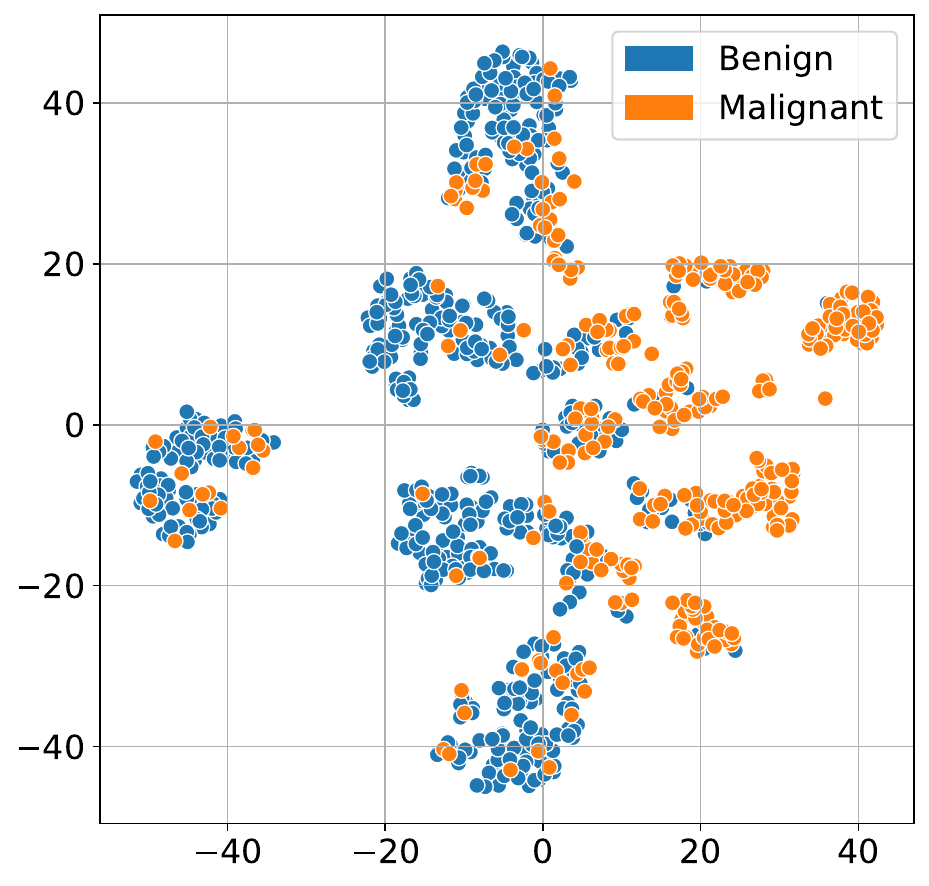}
    }
    \subfloat[Two-view features, $224^2$.]{
    \includegraphics[width=0.48\linewidth]{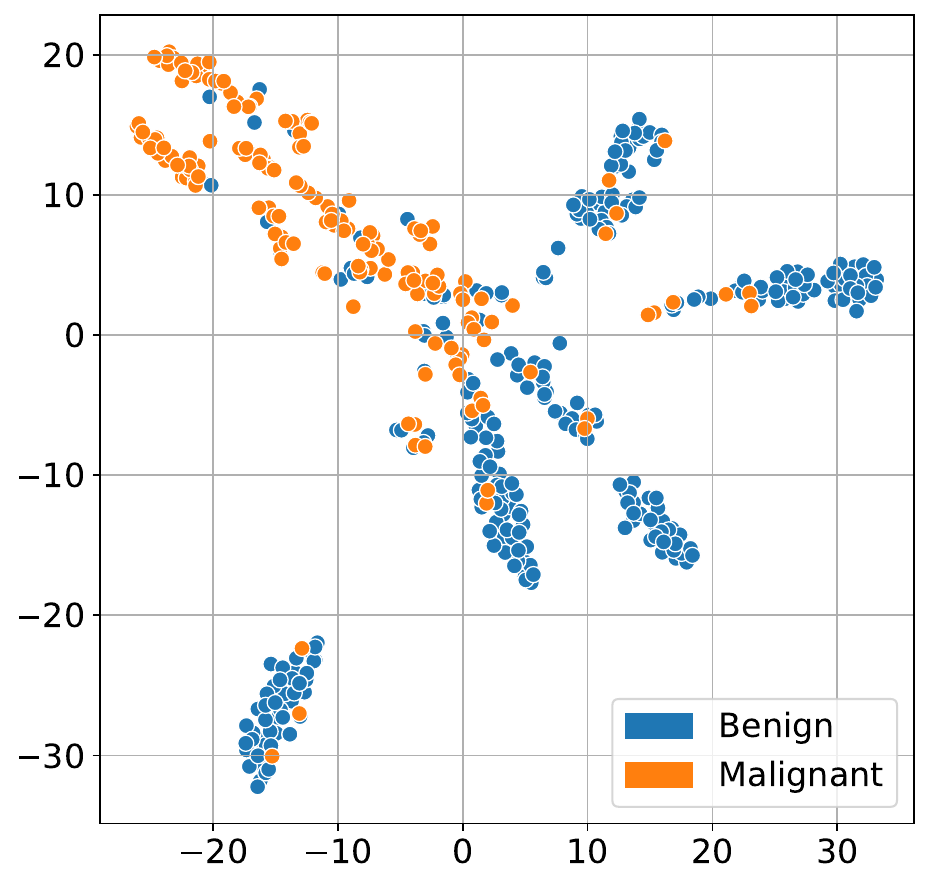}}
    \\
    \subfloat[Single-view features, $512^2$.]{
    \includegraphics[width=0.48\linewidth]{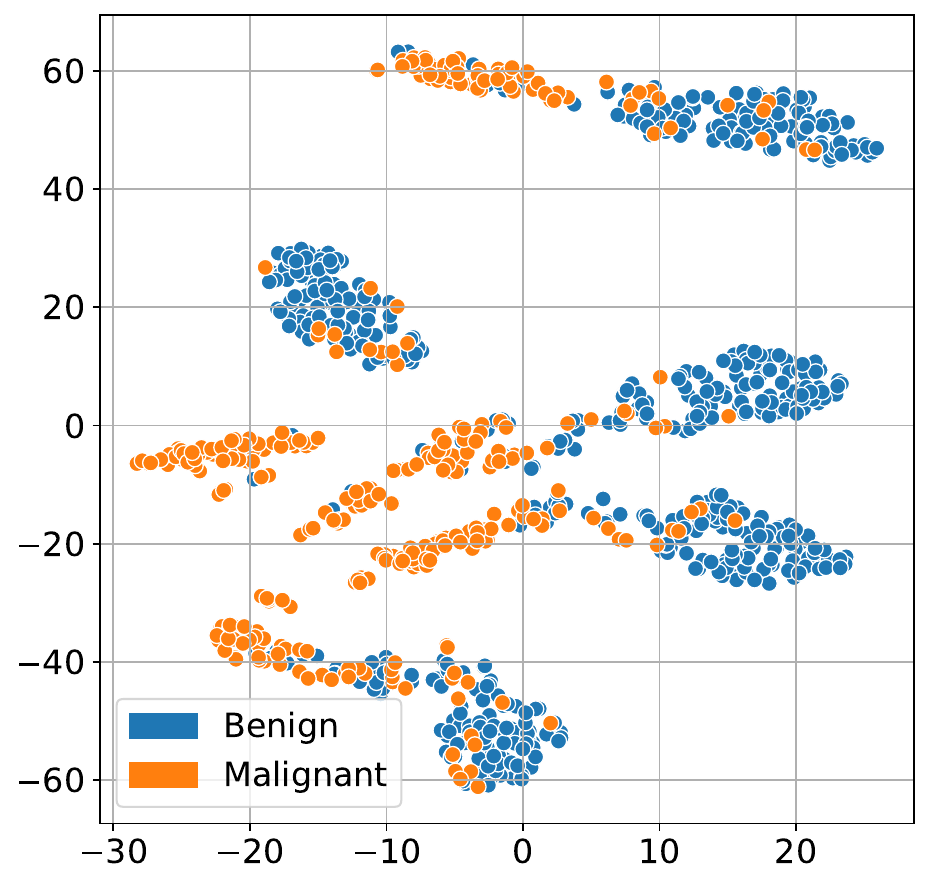}}
    \subfloat[Two-view features, $512^2$.]{
    \includegraphics[width=0.48\linewidth]{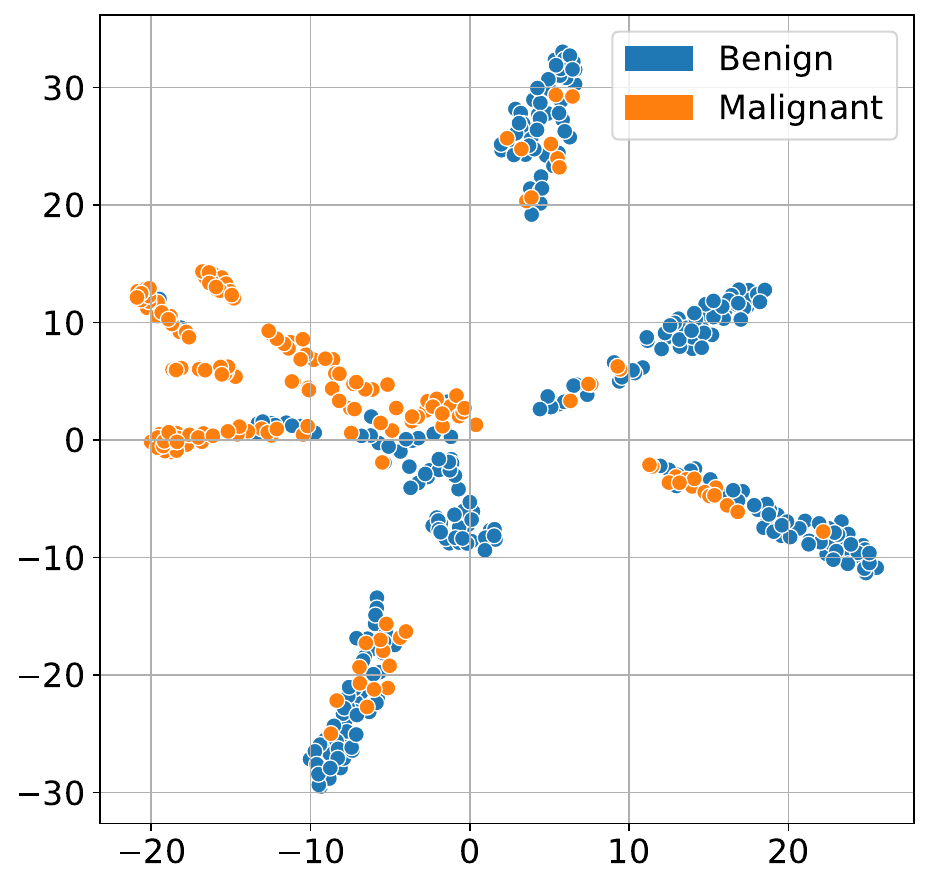}}\vspace{-5pt}
\caption{\textbf{t-SNE visualization for diagnosis task.} Higher resolution increases benign/malignant separation in the latent space.} 
\vspace{-15pt}
\label{fig:tsne}
\end{figure}

\noindent\textbf{Breast cancer diagnosis:}
Table~\ref{tab:pathology} shows that EfficientNet-B0 consistently outperformed ResNet-50 and DenseNet-121 in terms of AUC, indicating its strong capability for mammographic image analysis. For instance, single-view EfficientNet-B0 achieves 90.66\% AUC at $224^2$ and 92.13\% AUC at $512^2$, outperforming both ResNet-50 (87.42\% / 86.69\%) and DenseNet-121 (87.14\% / 90.52\%).  Higher input resolutions generally improve performance, highlighting the importance of preserving fine mammography details. Nevertheless, $224^2$ inputs still provide competitive results, offering computational efficiency. For example, single-view EfficientNet-B0 has 4.01M parameters and 0.41G FLOPs at $224^2$, compared to 4.01M parameters and 2.13G FLOPs at $512^2$. Two-view models outperform single-view models across all backbones, highlighting the value of combining CC and MLO views. Notably, two-view EfficientNet-B0 with $512^2$ inputs achieves the highest overall performance, reaching an AUC of 93.54\%, representing the most effective configuration in our benchmark. Fig.~\ref{fig:tsne} illustrates the t-SNE visualization, which shows how EfficientNet-B0 learned representations separate benign and malignant cases more clearly at higher input resolution.

\noindent\textbf{BI-RADS classification:}
The BI-RADS classification benchmarks are summarized in Table~\ref{tab:birads}. Across all experimental settings, EfficientNet-B0 achieved the highest AUCs (92.80\% in the two-class classification and 83.27\% in the three-class classification), followed by Swin-T, which consistently ranked second (91.44\% and 81.68\%).

\noindent\textbf{Density classification:}
As presented in Table~\ref{tab: density}, Swin-T always achieved the highest AUC across both input resolutions, demonstrating its effectiveness for multi-class density prediction. With $512^2$ inputs, Swin-T achieved a Macro-AUC of 89.14\%, outperforming ResNet-50 (87.06\%), DenseNet-121 (86.39\%), and EfficientNet-B0 (84.72\%). While higher resolutions generally improved the performance of CNN-based models, Swin-T obtained a slightly higher AUC (89.43\%) at $224^2$ resolution. We attribute this to the fact that the publicly available pretrained weights (from PyTorch Torchvision) were trained with $224^2$ inputs, and Transformer-based attention mechanisms are particularly sensitive to input scale.

\subsection{Ablation Study And Discussion}\label{sec:Ablation Study And Discussion}

\noindent\textbf{Shared vs. independent backbone:}
We compared two-view models using either shared or independent backbone weights. In the independent setting, the two EfficientNet-B0 branches are initialized and trained separately, while in the shared setting, both views share the same backbone parameters. Table~\ref{tab:pathology ablation} shows that although the computational cost between the two approaches is identical, sharing weights substantially reduces model size from 8.34M to 4.34M parameters without sacrificing performance. The shared-backbone EfficientNet-B0 achieved 92.99\% mean AUC at $224^2$ and 93.54\% at $512^2$, comparable to or slightly better than the independent-backbone configuration, which reached 93.54\% and 92.71\%. These results indicate that shared backbones are sufficient for this task, which offers a favorable trade-off between efficiency and AUC.

\begin{table}[tb]
    \caption{\textbf{Shared vs. independent two-view EfficientNet-B0 backbones.} Flops are 0.82G for $224^2$ and 4.27G under $512^2$.}\vspace{-5pt}
    \centering
    \setlength{\tabcolsep}{1pt}
    \begin{tabular}{l|cccccc}
    \toprule
         \textbf{Backbone}&\textbf{Input}&\textbf{Params}&\textbf{ACC(\%)}&\textbf{AUC(\%)}&\textbf{F1(\%)}\\
         \midrule
          Indepen-&$224^2$&8.34M&84.82$\pm$3.00 & 93.54$\pm$2.38 & 77.41$\pm$6.13\\
          dent&$512^2$ &8.34M&88.46$\pm$4.91 & 92.71$\pm$4.63 &  82.19$\pm$8.61\\\midrule

          Shared &$224^2$&4.34M&86.11$\pm$2.47 & 92.99$\pm$3.04 & 79.39$\pm$4.18\\
          &$512^2$ &4.34M&85.04$\pm$5.92 & 93.54$\pm$3.88 & 75.75$\pm$12.85\\
         \bottomrule
    \end{tabular}\vspace{-10pt}
    \label{tab:pathology ablation}
\end{table}

\begin{figure}[tb]
    \centering
    \subfloat[$224^2$ input.]{
    \includegraphics[width=.5\linewidth]{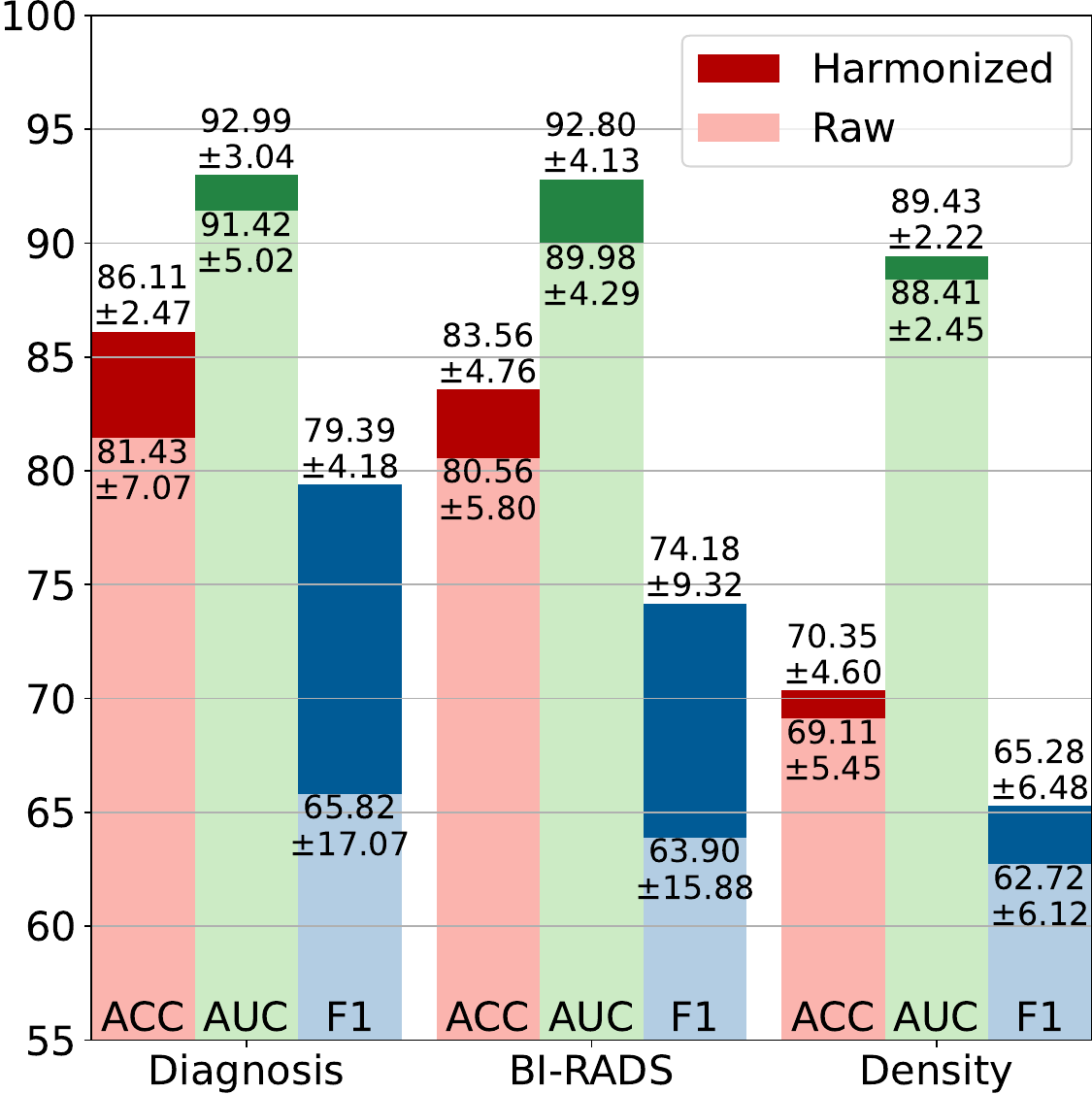}
    }
    \subfloat[$512^2$ input.]{
    \includegraphics[width=.5\linewidth]{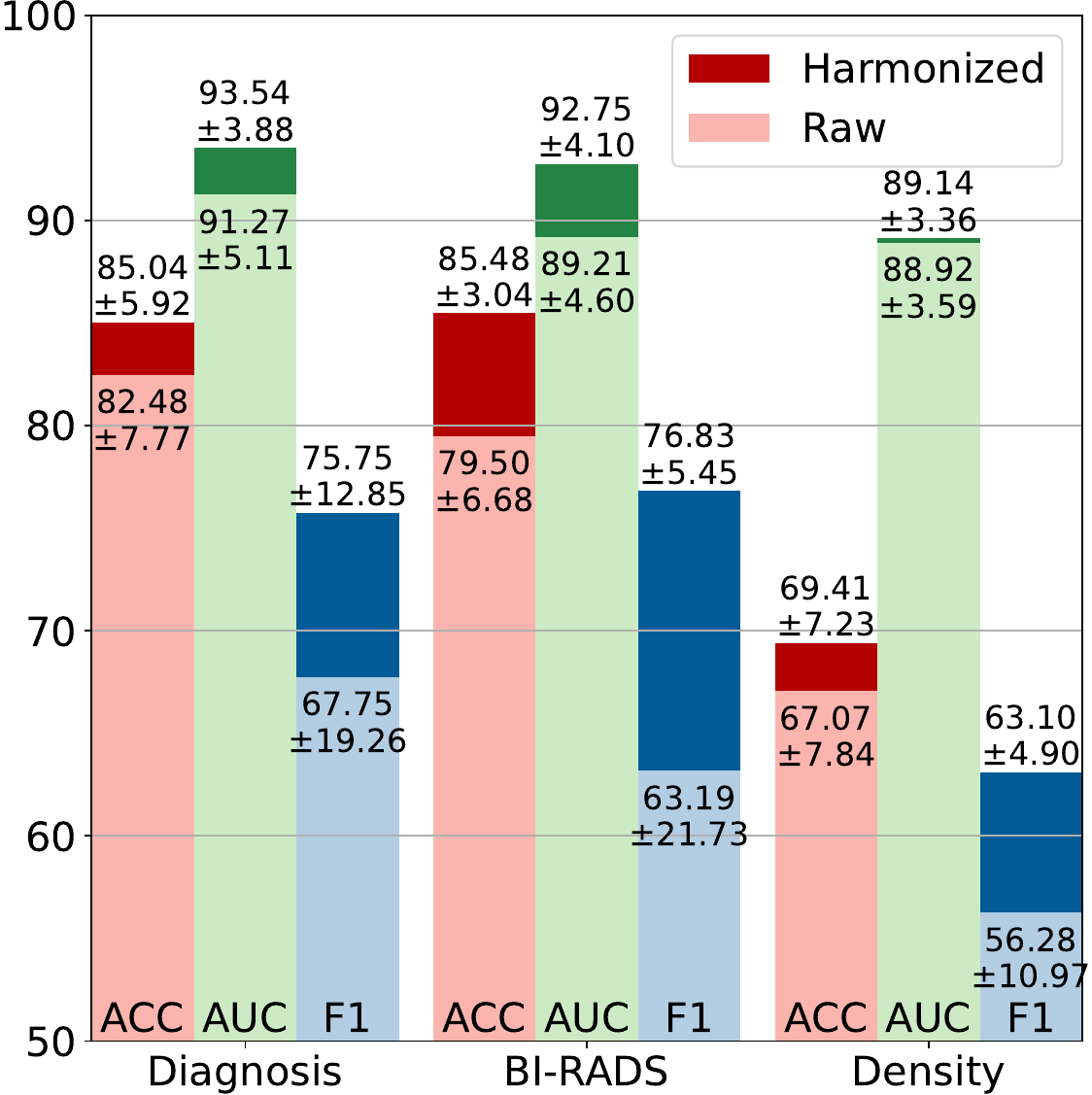}
    }\\
    \subfloat[Two-view diagnosis.]{
    \includegraphics[width=.5\linewidth]{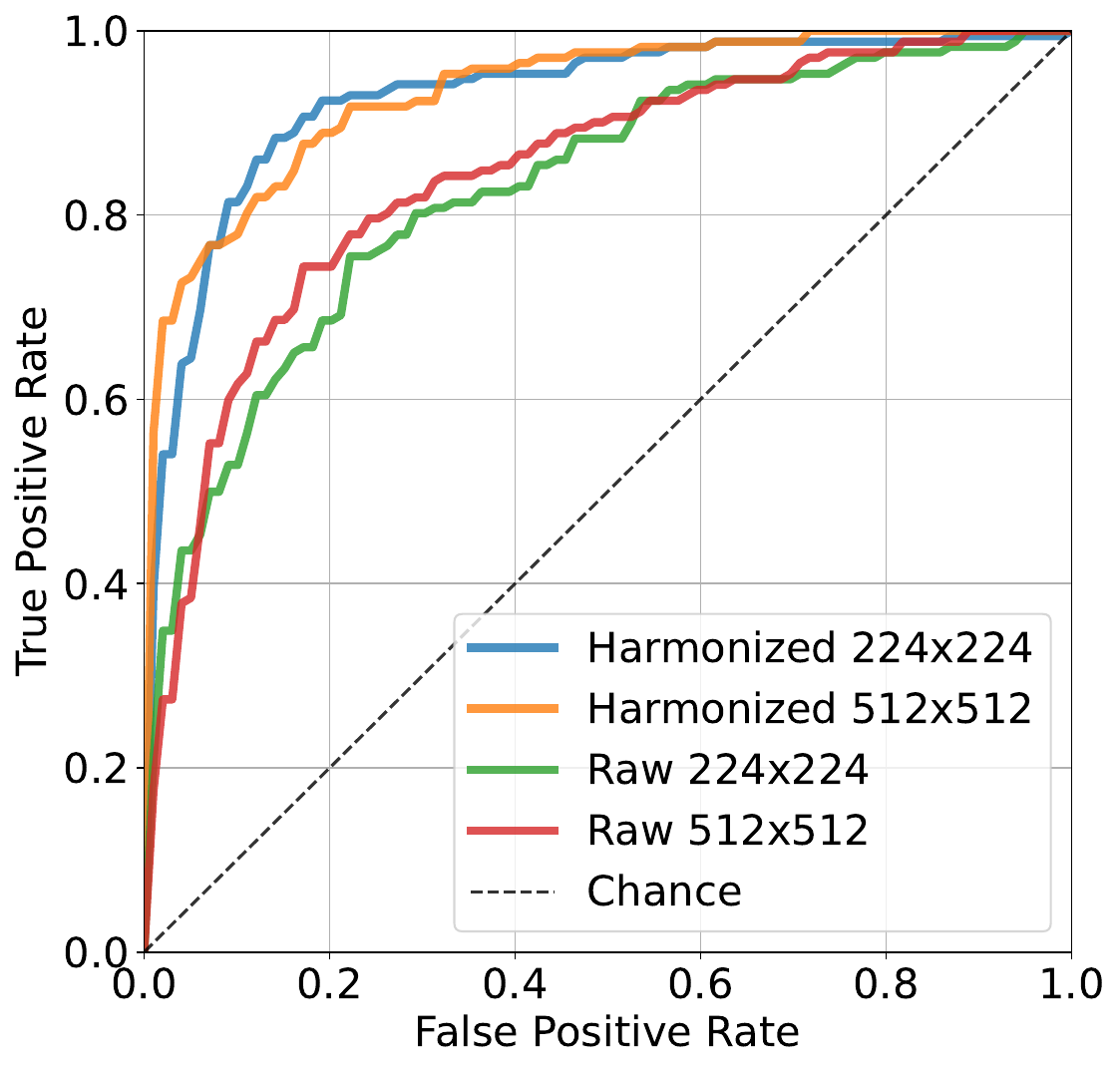}
    }
    \subfloat[Two-class BI-RADS.]{
    \includegraphics[width=.5\linewidth]{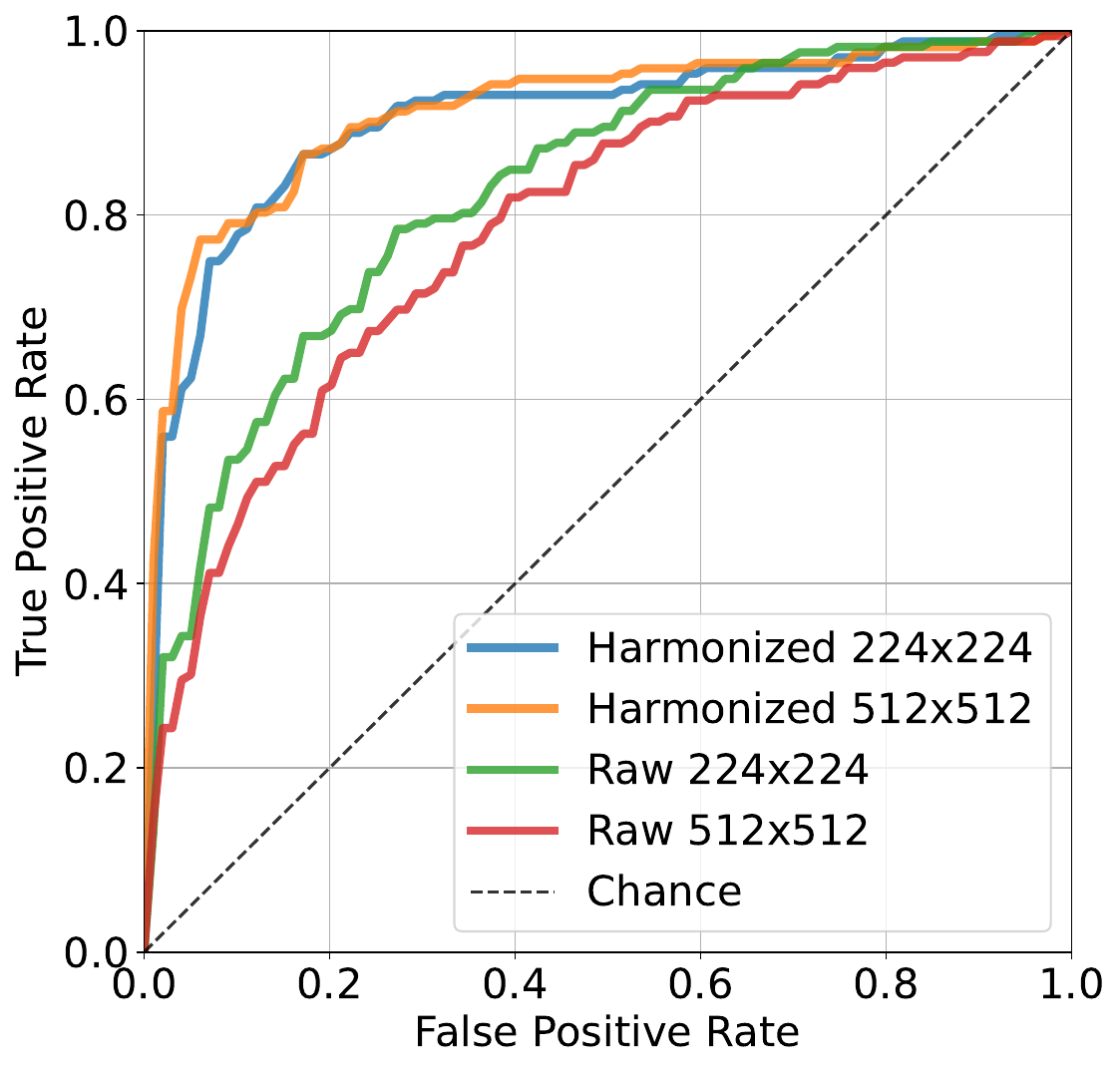}
    }
    \vspace{-5pt}
    \caption{\textbf{Harmonization improves ACC, AUC, and F1-Score across tasks.} Bars show mean$\pm$std over 5 folds, and curves show the mean ROC.
\textit{How to read:} In (a) and (b), darker bars (harmonized) consistently exceed lighter bars (raw). In (c) and (d), areas under the blue and orange (harmonized) curves are respectively larger than areas under the green and red (raw) curves.
\textit{Takeaway:} Pixel-space CDF alignment helps regardless of backbone or input size, supporting its use as a robust pre-processing step.}\vspace{-10pt}
    \label{fig:ablation}
\end{figure}

\noindent\textbf{Effect of foreground-only histogram harmonization:} We compared the performance of the best model in each task on raw mammograms versus harmonized mammograms. In this section, the conversion from MONOCHROME1 to MONOCHROME2 was still applied. As shown in Fig.~\ref{fig:ablation}, histogram matching consistently improved the performance across all settings. 
These results highlight that aligning intensity distributions across multi-vendor mammograms enhances the model’s discrimination between benign and malignant cases, making histogram matching an effective preprocessing step for robust pathology classification.

\begin{figure}[tb]
    \centering
    \subfloat[Malignant case 57L.\label{fig:gradcam1}]{
    \includegraphics[width=.7\linewidth,page=1]{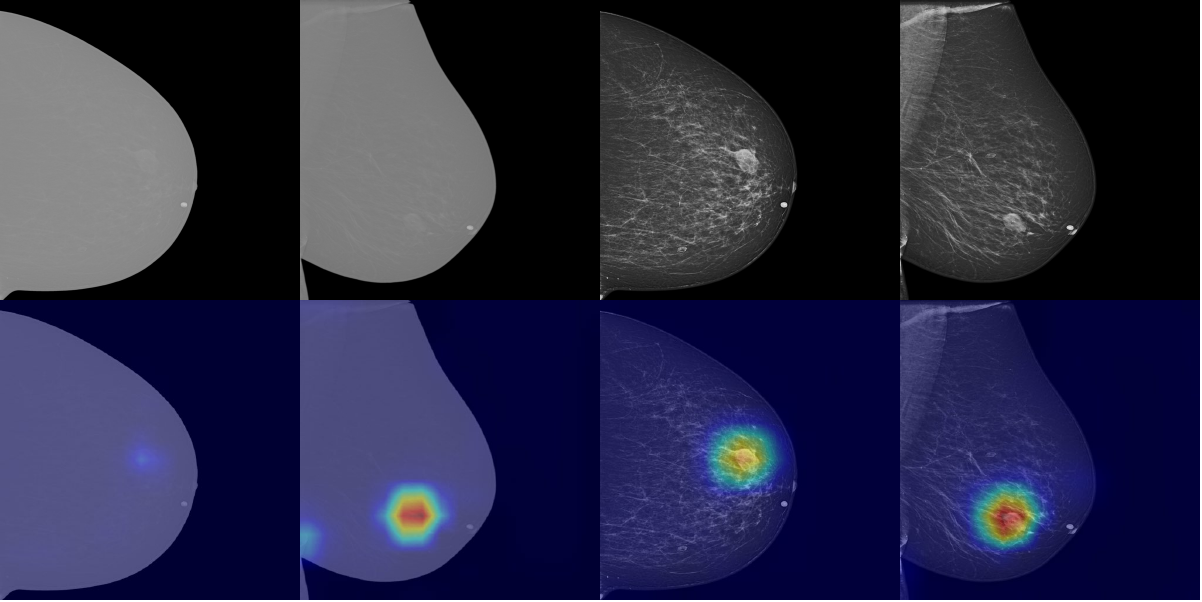}
    }\\
    \subfloat[Malignant case 83R.\label{fig:gradcam2}]{
    \includegraphics[width=.7\linewidth,page=2]{figures/gradcam.pdf}}\\
    \subfloat[Malignant case 105R.\label{fig:gradcam3}]{
    \includegraphics[width=.7\linewidth,page=3]{figures/gradcam.pdf}
    }\vspace{-5pt}
    \caption{\textbf{Attention becomes more focal after harmonization.}
Shown are raw (left two columns) vs. harmonized (right two columns) inputs with Grad-CAM overlays for two malignant cases. 
\textit{Observation:} Harmonization reduces diffuse activations and concentrates attention on lesion-bearing regions.
\textit{Implication:} Harmonization not only improves metrics (Fig.~\ref{fig:ablation}) but may also enhance clinical interpretability by focusing on suspicious tissue.}
   \vspace{-10pt} \label{fig:gradcam}
\end{figure}

Furthermore, we visually compared Grad-CAM attention maps~\cite{selvaraju2017grad} generated by the breast cancer diagnosis classifier on raw and histogram-harmonized two-view mammograms ($512^2$). As shown in Fig.~\ref{fig:gradcam}, harmonization improves the spatial focus of the model’s activations, directing attention toward clinically relevant regions. In Fig.~\ref{fig:gradcam1}, the model trained on raw images attended primarily to the MLO view, whereas the harmonized model successfully identified the suspicious area in both CC and MLO views. In Fig.~\ref{fig:gradcam2} and Fig.~\ref{fig:gradcam3}, the raw-image model’s attention was distracted by the chest wall above or below the breast, while harmonization guided the model toward the actual breast tissue, better aligning with lesion-related structures. 

\begin{figure}[tb]
   \centering
   \includegraphics[width=.8\linewidth]{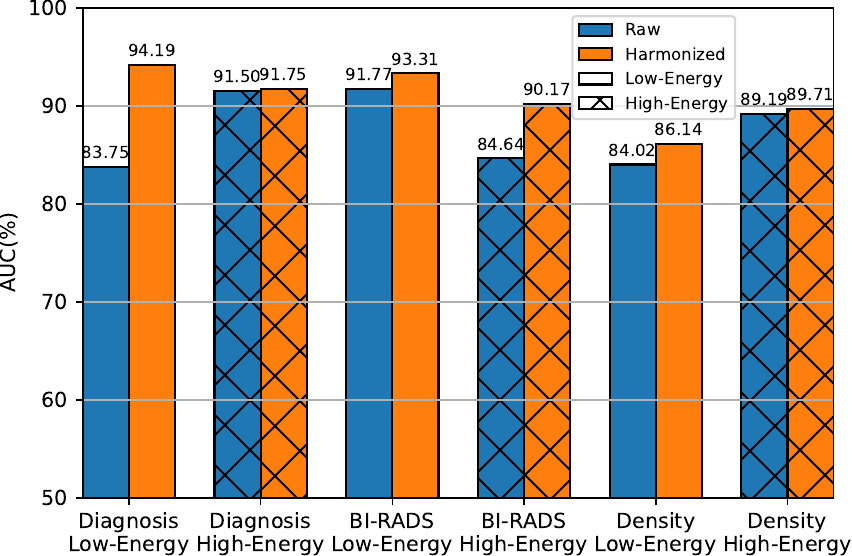}
   \vspace{-5pt}
   \caption{\textbf{Energy-specific AUC.} Histogram harmonization improved AUC, particularly for low-energy images.}
   \vspace{-15pt}
   \label{fig:energy}
\end{figure}

\noindent\textbf{Energy-specific analysis:} 
To evaluate energy-specific performance, predictions and labels from five patient-specific folds were concatenated to obtain full-dataset predictions. AUC was then computed separately for low- and high-energy subsets using the best models for each task (EfficientNet-B0 for two-view diagnosis and two-class BI-RADS, and Swin-T for density). All models were with $224^2$ inputs. As shown in Fig.~\ref{fig:energy}, the histogram harmonization reduces the distribution gap between the two energy types, improving generalization. Notably, when high-energy images dominate the dataset, low-energy images benefit most from harmonization.

\section{Conclusion}
We present LUMINA, a multi-vendor FFDM dataset with comprehensive pathology, BI-RADS, and density annotations. We also introduced a foreground-only histogram-based harmonization method to mitigate inter-scanner variability. Extensive experiments demonstrate that harmonization consistently improves performance across multiple tasks and architectures. These findings highlight the importance of vendor-level normalization for robust mammography AI. We expect LUMINA to bridge the gap between academic benchmarks and real-world clinical deployment.


\section{Data and Code}
The LUMINA dataset and the associated source code are publicly available at the following links:
\begin{itemize}
    \item \textbf{OSF (Dataset):} \url{https://osf.io/b63jc/}
    \item \textbf{Kaggle (Dataset):} \url{https://www.kaggle.com/datasets/phy710/lumina-mammography-dataset}
    \item \textbf{GitHub (Code):} \url{https://github.com/NUBagciLab/LUMINA}
\end{itemize}

\section{Acknowledgment}
This research was partially funded by NIH: R01-HL171376. 

{
    \small
    \bibliographystyle{ieeenat_fullname}
    \bibliography{main}
}


\clearpage
\setcounter{page}{1}
\maketitlesupplementary
\begin{figure}[tb]
    \centering
    \subfloat[Single-view $224^2$ harmonized.]{
    \includegraphics[width=0.48\linewidth]{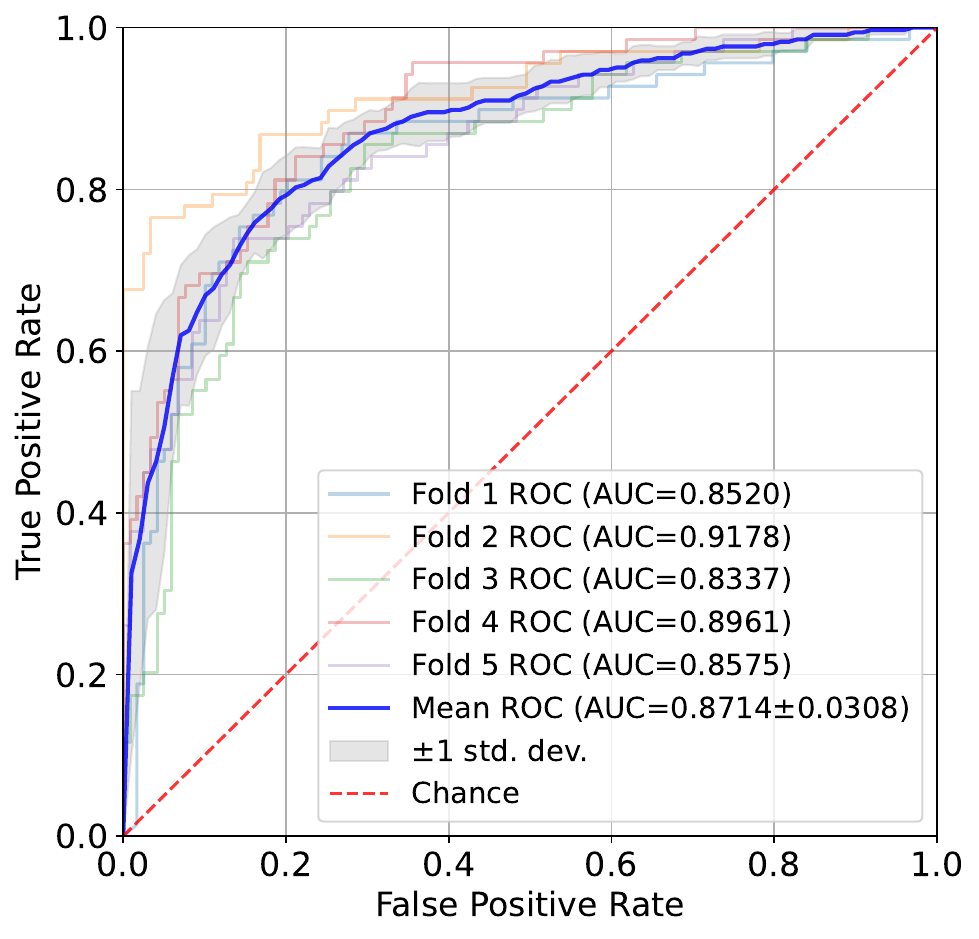}
    }
    \subfloat[Single-view $512^2$ harmonized.]{
    \includegraphics[width=0.48\linewidth]{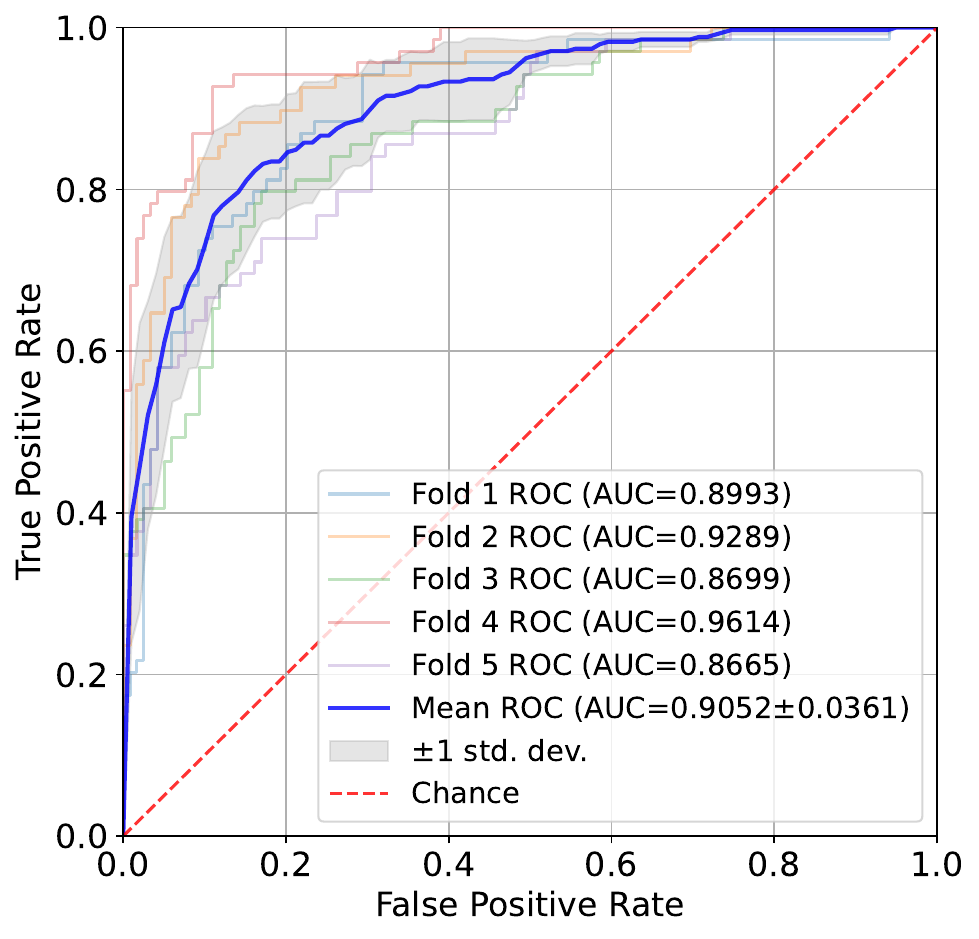}
    }\\
    \subfloat[Two-view $224^2$  harmonized.]{
    \includegraphics[width=0.48\linewidth]{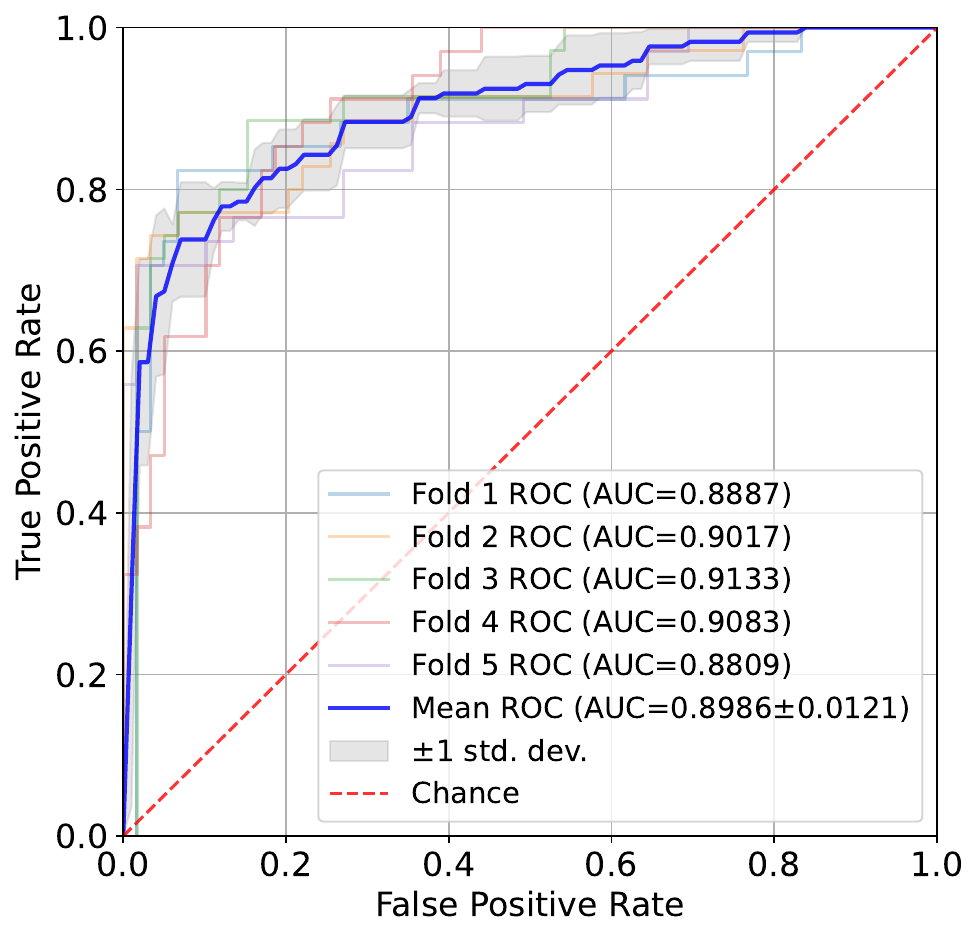}
    }
    \subfloat[Two-view $512^2$ harmonized.]{
    \includegraphics[width=0.48\linewidth]{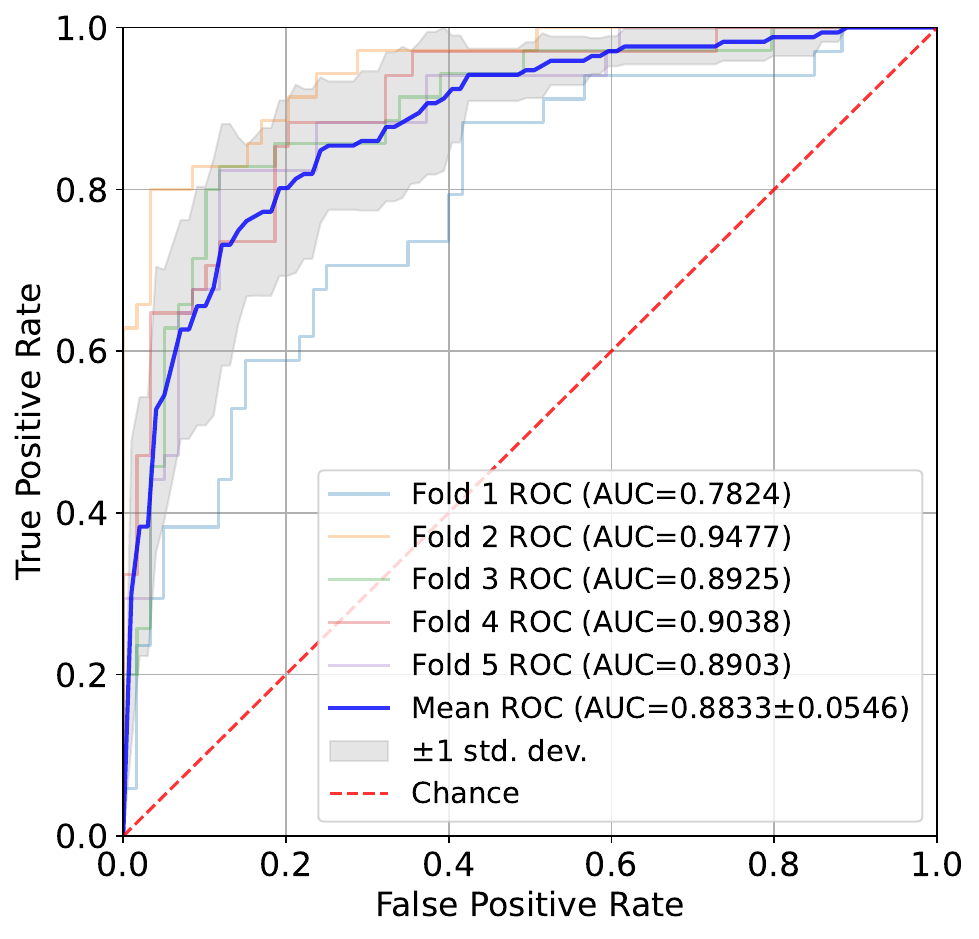}
    }\\
    \subfloat[Two-view $224^2$ raw input.]{
    \includegraphics[width=0.48\linewidth]{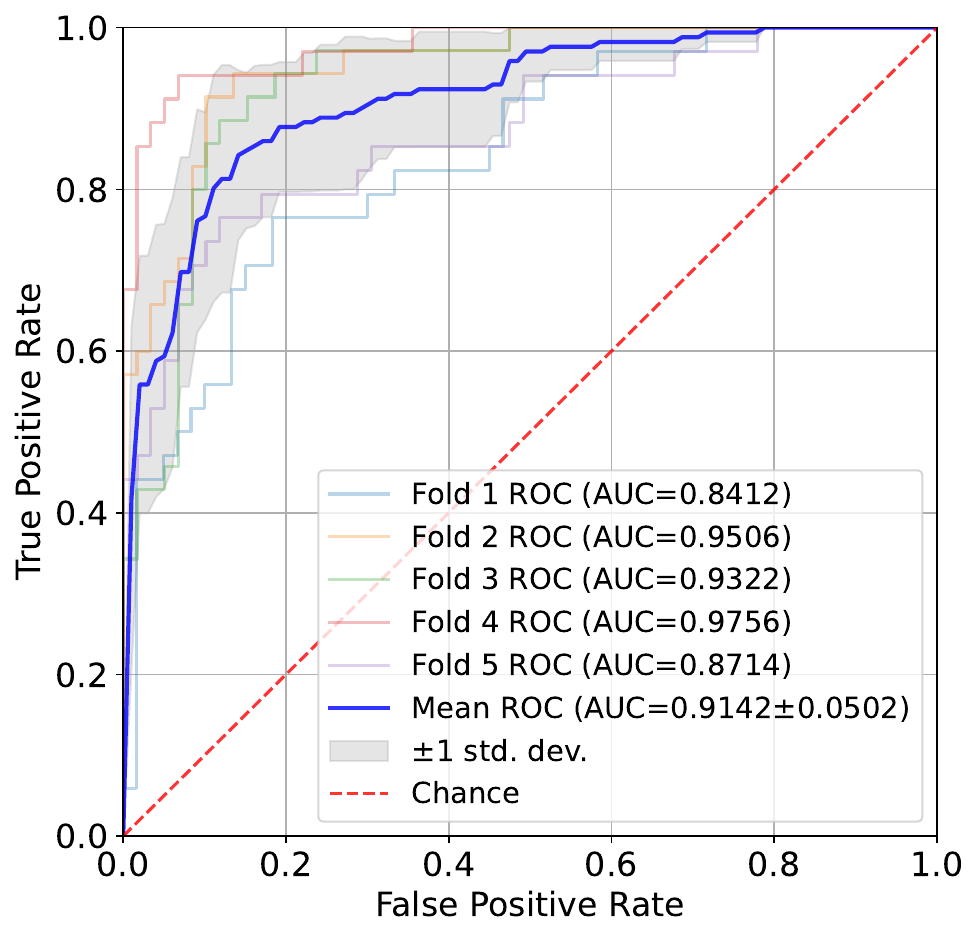}
    }
    \subfloat[Two-view $512^2$ raw input.]{
    \includegraphics[width=0.48\linewidth]{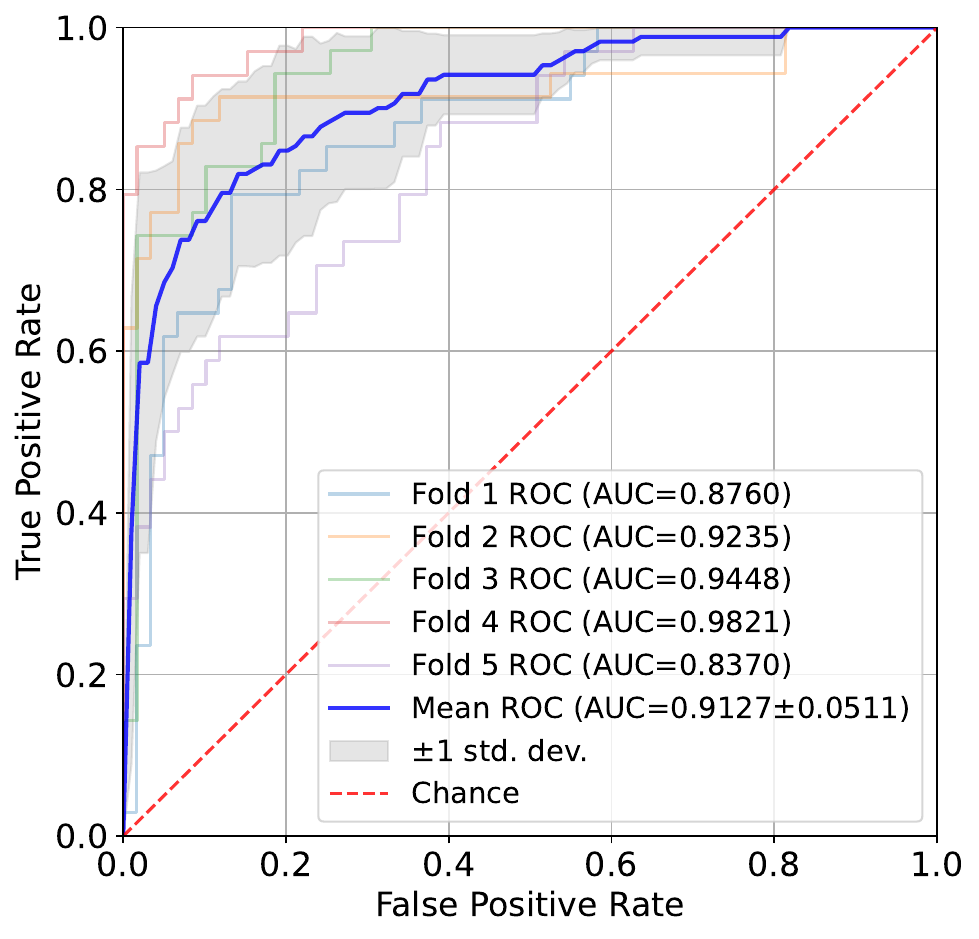}
    }
    \caption{\textbf{ROC curves of the EfficientNet-B0 model for breast cancer diagnosis.}}
    \label{fig:roc diagnosis}
\end{figure}

\begin{figure}[tb]
    \centering
    \subfloat[$224^2$ harmonized input.]{
    \includegraphics[width=0.48\linewidth]{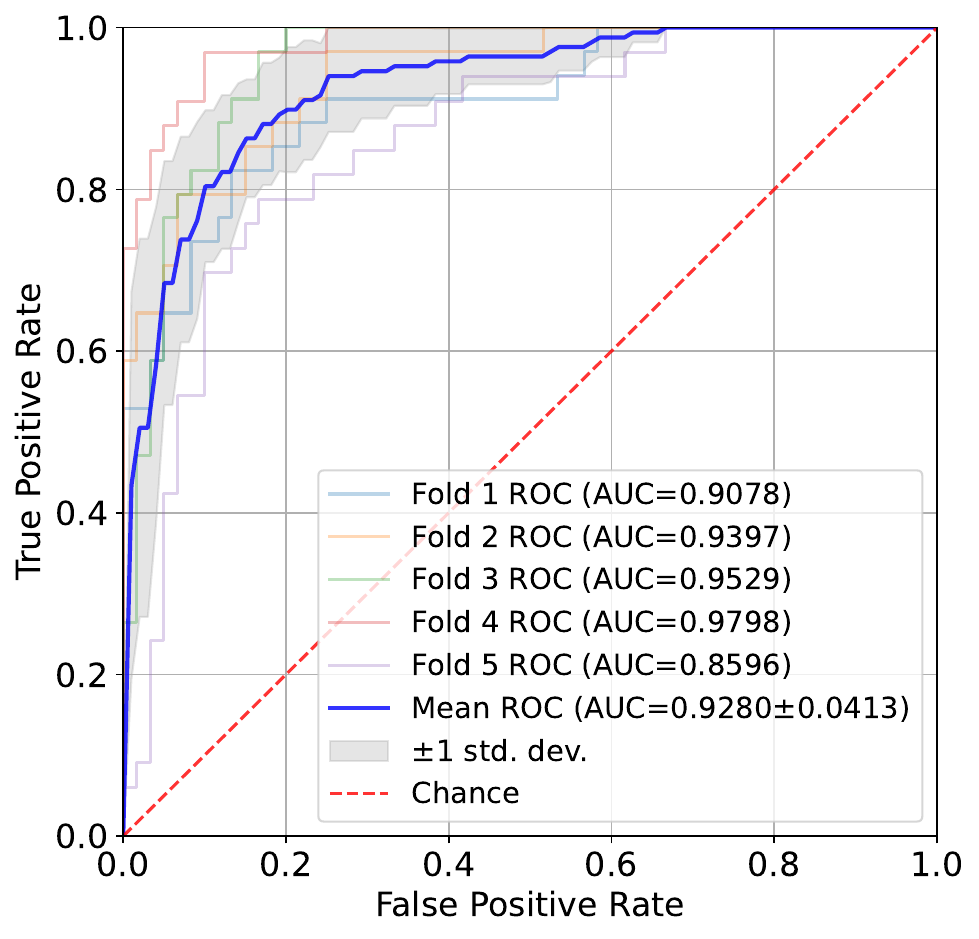}
    }
    \subfloat[$512^2$ harmonized input.]{
    \includegraphics[width=0.48\linewidth]{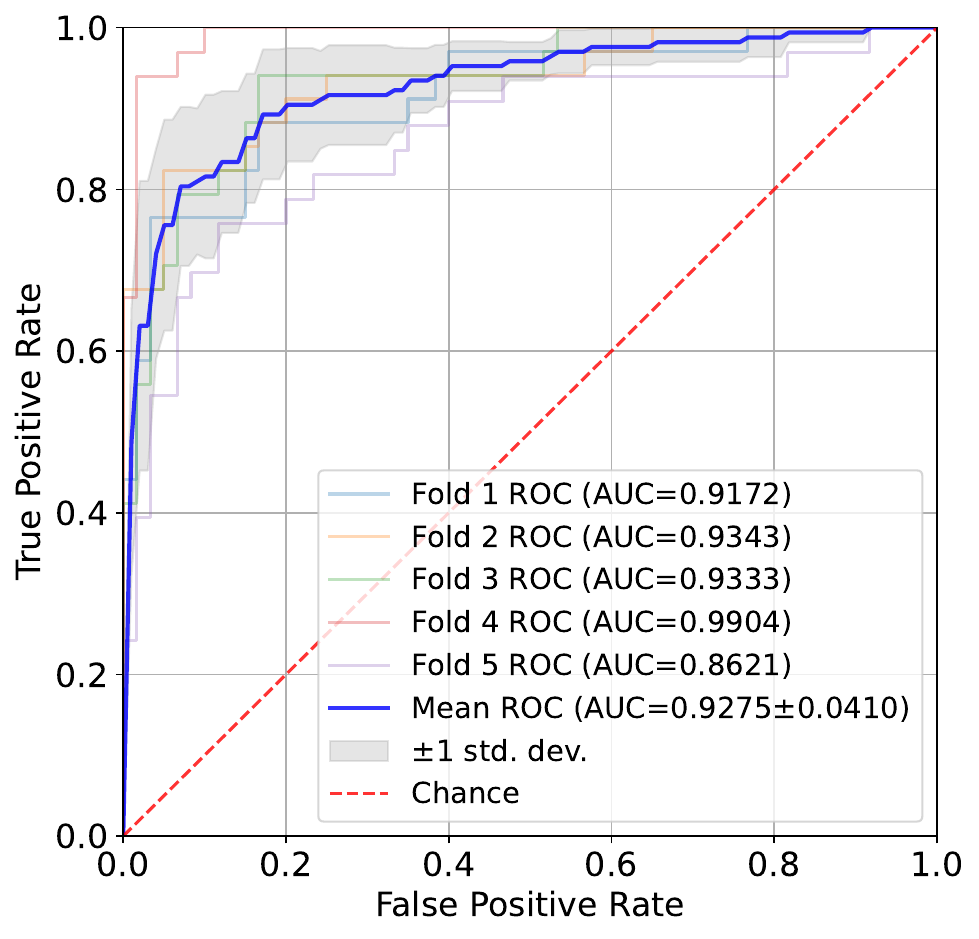}
    }\\
    \subfloat[$224^2$ raw input.]{
    \includegraphics[width=0.48\linewidth]{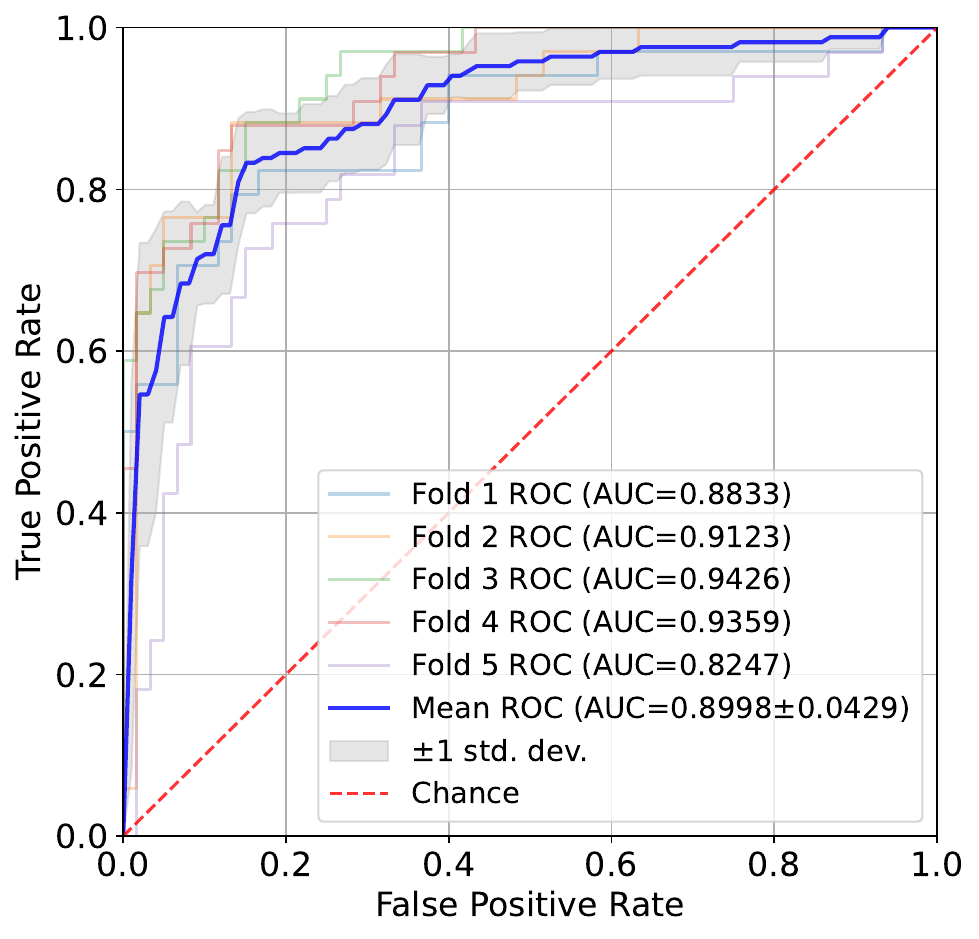}
    }
    \subfloat[$512^2$ raw input.]{
    \includegraphics[width=0.48\linewidth]{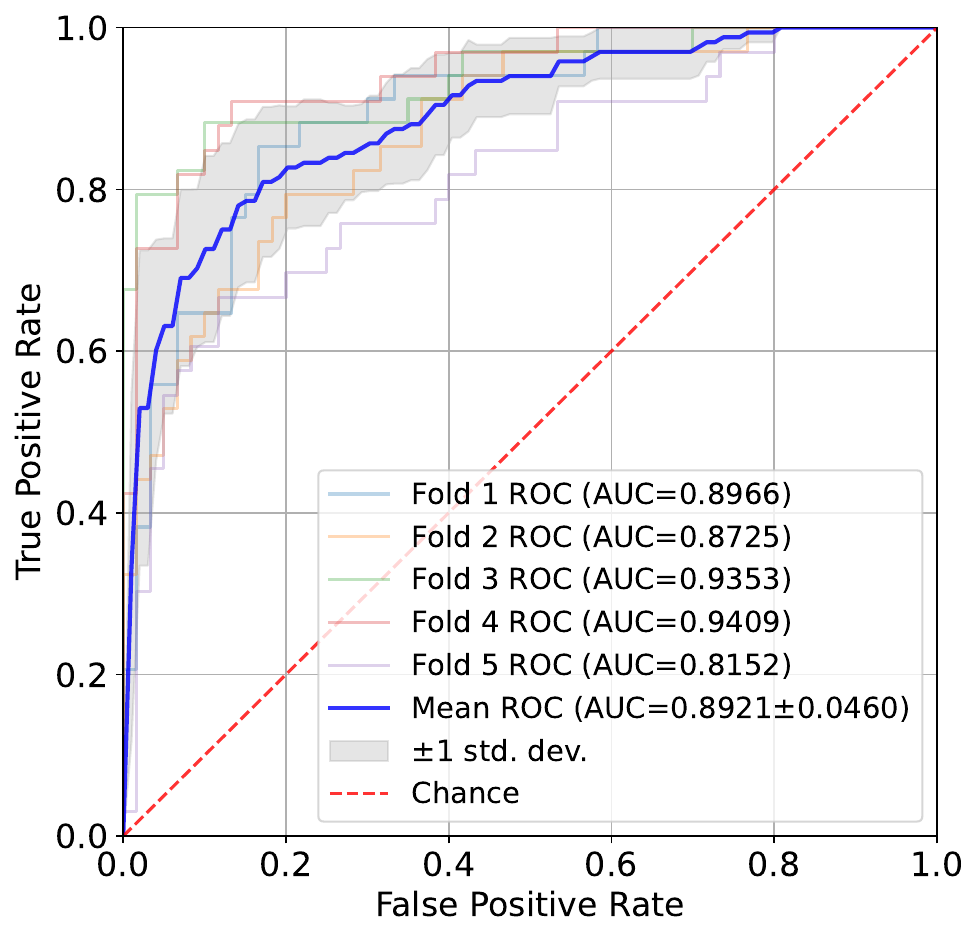}
    }
    \caption{\textbf{ROC curves of the EfficientNet-B0 model for two-class BI-RADS classification.}}
    \label{fig:roc birads}
\end{figure}

Fig.~\ref{fig:roc diagnosis} shows the ROC curves of EfficientNet-B0 for the breast cancer diagnosis task, where we plot the ROC for each fold and the mean ROC. The two-view model consistently outperforms the single-view variant, demonstrating the benefit of combining CC and MLO information. Higher input resolution ($512^2$) further improves AUC compared to the $224^2$ setting. In all configurations, object-histogram-based harmonization yields a noticeable upward shift in the ROC curves, confirming its effectiveness in stabilizing image appearance and enhancing diagnostic performance. Fig.~\ref{fig:roc birads} shows the ROC curves of EfficientNet-B0 for the two-class BI-RADS classification task. It also supports that higher resolution and harmonization improves the results.



Table~\ref{tab:Positioning} lists the distinguishing characteristics of LUMINA compared to existing mammography datasets and prior harmonization strategies. 

Table~\ref{tab:hyperparity} summarizes the unified hyperparameter settings and pre-processing operations used across all model architectures, respectively. These settings and operations ensured a controlled and fair comparison in our benchmark experiments.

\begin{table*}[htbp]
\centering
\small
\caption{\textbf{Positioning.} LUMINA provides pathology, BI-RADS, and density on multi-vendor FFDM, plus a simple, effective pixel‑space harmonization baseline.}
\begin{tabular}{p{1.8cm} p{3.0cm} p{2.8cm} p{3.8cm} p{3.8cm}}
\toprule
\textbf{Family} & \textbf{Examples} & \textbf{Labels/Supervision} & \textbf{Architecture/Objective} & \textbf{Limitations vs.\ LUMINA} \\
\midrule
\textbf{Film (SFM) datasets} & MIAS~\cite{suckling1994mammographic}, DDSM~\cite{bowyer1996digital}, CBIS-DDSM~\cite{lee2017curated} & Pathology (yes), BI-RADS (varies) & Early CNNs / CAD; single-task classification & Film scans, lower resolution; limited modern FFDM relevance. \\
\midrule
\textbf{Digital (FFDM) datasets} & INbreast~\cite{moreira2012inbreast}, VinDR-Mammo~\cite{pham2022vindr}, RSNA~\cite{carr2022rsna}, CMMD~\cite{cai2023online}, KAU-BCMD~\cite{alsolami2021king} & Pathology or BI-RADS; density varies & CNNs/ViTs; single- or dual-task & Often single vendor/system; partial labels; limited multi-task scope. \\
\midrule
\textbf{Harmonization (feature space)} & ComBat~\cite{orlhac2022guide} & Batch-effect correction on features & Empirical Bayes (location/scale) on radiomics/latent features & Requires feature extraction; not pixel-space; modality-agnostic assumptions. \\
\midrule
\textbf{Harmonization (federated learning)} & HarmoFL~\cite{jiang2022harmofl} & Federated frequency‑domain drift normalization & Joint optimization across clients & Heavy infra; task/model coupling; not a simple preproc. \\
\midrule
\textbf{LUMINA (Ours)} & Multi-vendor FFDM (6 systems); energy metadata & Pathology + BI-RADS + density & \emph{Foreground-only} pixel-space CDF matching + unified 3‑task benchmark; single-/two-/four‑view baselines & Vendor/energy diversity; three tasks; consistent harmonization gains; improved attention focus. \\
\bottomrule
\end{tabular}\label{tab:Positioning}
\end{table*}

\begin{table*}[htbp]
\centering
\small
\setlength{\tabcolsep}{5pt}
\caption{\textbf{Hyperparameter parity across backbones.} Shared schedule and augmentations ensure a fair comparison (see Sec.~\ref{sec:Experimental Setup}).}
\begin{tabular}{lccccccc}
\toprule
\textbf{Backbone} & \textbf{Pre-trained} & \textbf{Learning Rate} & \textbf{Weight Decay} & \textbf{Epochs} & \textbf{Learning Rate scheduler} & \textbf{Batch} & \textbf{Augmentation} \\
\midrule
ResNet-50 & ImageNet-1K & $1\!\times\!10^{-3}$ & $1\!\times\!10^{-5}$ & 100 & $\times 0.1$ every 30 & 32 & flip, resize \\
DenseNet-121 & ImageNet-1K & $1\!\times\!10^{-3}$ & $1\!\times\!10^{-5}$ & 100 & $\times 0.1$ every 30 & 32 & flip, resize \\
EfficientNet-B0 & ImageNet-1K & $1\!\times\!10^{-3}$ & $1\!\times\!10^{-5}$ & 100 & $\times 0.1$ every 30 & 32 & flip, resize \\
Swin-T & ImageNet-1K & $1\!\times\!10^{-5}$ & $1\!\times\!10^{-5}$ & 100 & $\times 0.1$ every 30 & 32 & flip, resize \\
\bottomrule
\end{tabular}
\label{tab:hyperparity}
\end{table*}

\begin{table}[htbp]
\centering
\small
\caption{\textbf{Shared pre-processing.} Ensures identical inputs across backbones; isolates model differences.}
\begin{tabular}{lp{5.6cm}}
\toprule
\textbf{Stage} & \textbf{Operation (applied to all models equally)} \\
\midrule
DICOM handling & Convert MONOCHROME1 to MONOCHROME2; remove text burn-ins. \\
Foreground mask & Define $M=\{(x,y)\mid \mathbf{I}(x,y)>0\}$; exclude background from histograms. \\
Harmonization & Foreground CDF matching to low-energy reference (Eqs.~1--8); training-free, model-agnostic. \\
Resize/replicate & Resize to $224^2$ or $512^2$; replicate grayscale to 3 channels for ImageNet backbones (Sec.~\ref{sec:Experimental Setup}). \\
Augment & Random horizontal flip (all backbones, all tasks). \\
Normalization & As in backbone defaults; no per-model special tuning beyond LR noted in Sec.~\ref{sec:Experimental Setup}. \\
\bottomrule
\end{tabular}\label{tab:preprocessing}
\end{table}

\end{document}